\documentclass[]{mn2e}
\usepackage{mathrsfs,amssymb}
\usepackage{mathtools}
\usepackage{amsmath}
\usepackage{graphicx}
\usepackage{tabularx}
\usepackage{epsfig}
\usepackage{rotating}
\usepackage{diagbox}
\usepackage{pdflscape}
\usepackage{bm}
\newcommand {\Ks}  {{\rm K_S}}
\newcommand {\AV}  {A_{V}}
\newcommand {\AKs} {{A_{\Ks}}}

\newcommand {\lambdapeak}{\lambda_{\rm peak}}
\newcommand {\gammasil}{\gamma_{\rm sil}}
\newcommand {\RV}{R_V}

\newcommand {\mum}{\,{\rm \mu m}}
\newcommand       \K            {\,{\rm K}}
\newcommand       \km           {\,{\rm km}}
\newcommand       \s            {\,{\rm s}}
\newcommand	\beq	{\begin{equation}}	%{\begin{displaymath}}
\newcommand	\eeq	{\end{equation}}	%{\end{displaymath}}
\newcommand       \Angstrom     {\,{\rm \AA}}
\newcommand       \simali       {\sim\,}
\newcommand       \magni        {\,{\rm mag}}
\newcommand       \simlt        {\lesssim}
\newcommand       \simgt        {\gtrsim}
\newcommand       \gtsim        {\gtrsim}

\newcommand       \Teff         {T_{\star}}
\title{
Probing the 9.7$\mum$ Interstellar Silicate
Extinction Profile through the \emph{Spitzer}/IRS
Spectroscopy of OB Stars
}

\author[Shao, Jiang, Li, Gao, Lv \& Yao]
       {Zhenzhen Shao$^{1,2,3}$\thanks{zhenzhenshao@mail.bnu.edu.cn},
        B.W.~Jiang$^{1}$\thanks{bjiang@bnu.edu.cn},
        Aigen Li$^{2}$\thanks{lia@missouri.edu},
        Jian Gao$^{1}$,
        Zhangpan Lv$^{1}$, and
        Jiawen Yao$^{1}$\\
        $^1$Department of Astronomy,
            Beijing Normal University,
            Beijing 100875, China\\
        $^2$Department of Physics and Astronomy,
             University of Missouri,
             Columbia, MO 65211, USA\\
        $^3$Beijing Ancient Observatory,
            Beijing Planetarium,
            Beijing 100005, China
                 }
\begin{document}
\date{Received date  / Accepted date }
\pagerange{\pageref{firstpage}--\pageref{lastpage}} \pubyear{2018}

\maketitle

\label{firstpage}
\begin{abstract}
The 9.7$\mum$ interstellar spectral feature,
arising from the Si--O stretch of amorphous silicate dust,
is the strongest extinction feature in the infrared (IR).
In principle,  the spectral profile of this feature
could allow one to diagnose the mineralogical
composition of interstellar silicate material.
However, observationally, the 9.7$\mum$ interstellar
silicate extinction profile is not well determined.
Here we utilize the \emph{Spitzer}/IRS spectra
of five early-type (one O- and four B-type) stars
and compare them with that of unreddened stars
of the same spectral type to probe the interstellar
extinction of silicate dust around $9.7\mum$.
We find that, while the silicate extinction profiles
all peak at $\simali$9.7$\mum$, two stars exhibit
a narrow feature of FWHM\,$\simali$2.0$\mum$
and three stars display a broad feature of
FWHM\,$\simali$3.0$\mum$.
We also find that the width
of the 9.7$\mum$ extinction feature
does not show any environmental dependence.
With a FWHM of $\simali$2.2$\mum$,
the mean 9.7$\mum$ extinction profile,
obtained by averaging over our five stars,
closely resembles that of the prototypical
diffuse interstellar medium
along the lines of sight toward
Cyg OB2 No.\,12 and WR\,98a.
Finally, an analytical formula is presented
to parameterize the interstellar extinction
in the IR at $0.9\mum\simlt\lambda\simlt15\mum$.
%from the  far ultraviolet to the mid-IR.
%
\end{abstract}


% \linenumbers

%% main text
\section{Introduction\label{sec:intro}}
The wavelength dependence of the interstellar extinction,
known as the interstellar extinction curve
or the interstellar extinction law,
varies substantially in the ultraviolet (UV) and
optical wavelengths (see Draine 2003).
Cardelli et al.\ (1989; hereafter CCM) found that
this variation can be closely parameterized by
a single parameter $R_V$,
where the dimensionless quantity
$R_V\equiv A_V/\left(A_B-A_V\right)$
is the total-to-selective extinction ratio
between the $B$ band
at $\lambda\approx4400\Angstrom$
and the $V$ band
at $\lambda\approx5500\Angstrom$.
As $R_V$ varies from one region to another,
the slope of the UV/optical extinction curve
shows considerable variations from being
steep in diffuse regions characterized by
a smaller $R_V$ (typically $\simlt3$) to
being flat in regions which are more dense
and characterized by a larger $R_V$
(typically $\simgt4$; see Li et al.\ 2015).
The 2175$\Angstrom$ extinction bump,
the strongest spectral feature
in the UV extinction curve,
also varies with $R_V$.
While the peak wavelength of
the 2175$\Angstrom$ bump is invariant,
its width ($\gamma$) increases with $R_V$
(see Figure~3.8 of Whittet 2003)
and its height ($\propto\gamma^{-2}$),
the excess extinction of the bump
over the underlying linear background extinction,
decreases with $R_V$ (see Figure~7 of CCM).
In contrast, in the near- and mid-infrared
wavelength range of $0.9\mum\simlt\lambda\simlt8\mum$,
the interstellar extinction curve
appears to be remarkably uniform over the sky
and shows little variation with $R_V$
(e.g., see Martin \& Whittet 1990,
Wang \& Jiang 2014, Xue et al.\ 2016).

At $\lambda\gtsim8\mum$, the extinction curve
exhibits a pronounced spectral feature
at $\simali$9.7$\mum$,
the strongest extinction feature in the IR.
This feature is attributed to the Si--O
stretching absorption of silicate dust (Henning 2010).
In principle, the spectral profile of
the 9.7$\mum$ silicate extinction feature
allows one to diagnose the mineralogical
composition of interstellar silicate material.\footnote{%
  For example, based on the absence of
  any substructures in the interstellar
  9.7$\mum$ feature which is broad,
  smooth, and featureless, Li \& Draine (2001),
  Kemper et al.\ (2004), Li et al.\ (2008),
  and Poteet et al.\ (2015) concluded that
  interstellar silicates are predominantly
  amorphous rather than crystalline.
  }
However, observationally, the spectral profile of
the 9.7$\mum$ silicate extinction is not as well
determined as the UV/optical extinction curve,
partly due to the substantial drop in extinction
from the UV, optical to the IR.
In the literature, one often relies on Mie theory
together with the optical constants synthesized
for ``astronomical silicates''
(Draine \& Lee 1984) or the experimentally-measured
optical constants of cosmic silicate analogues
(e.g., see J\"ager et al.\ 1994,
Dorschner et al.\ 1995) to calculate
the 9.7$\mum$ extinction profile from
submicron-sized spherical silicate grains.\footnote{%
  For submicron-sized silicate grains,
  the calculated 9.7$\mum$ extinction profile
  is insensitive to the exact grain size
  since they are in the Rayleigh regime
  (Bohren \& Huffman 1983).
  }
The optical constants of ``astronomical silicates''
at $\simali$8--13$\mum$ were constructed
based on the Trapezium emission profile
(Gillett et al.\ 1975) which peaks at
$\simali$9.56$\mum$ and has
a full width of half maximum (FWHM)
of $\simali$3.45$\mum$
(Draine \& Lee 1984).\footnote{%
  The Trapezium emissivity profile
  was chosen because it appears to
  provide a good fit to
  the then 9.7$\mum$ absorption band
  observed in dark clouds
  as well as emission by hot circumstellar dust
  around many oxygen-rich stars.
  Draine \& Lee (1984) had already recognized
  that some stars with circumstellar emission
  do show a narrower 9.7$\mum$ emission feature,
  e.g., the circumstellar 9.7$\mum$ emission
   feature of $\mu$ Cephei (Roche \& Aitken 1984),
   a red supergiant, peaks at $\simali$9.69$\mum$
   and has a width of $\simali$2.35$\mum$.
   }
In contrast, the peak wavelength and FWHM of
the 9.7$\mum$ silicate extinction feature
calculated from those experimentally-measured
optical constants vary with the silicate
composition (e.g., pyroxene vs. olivine)
and the iron content (e.g., see J\"ager et al.\ 1994,
Dorschner et al.\ 1995).\footnote{%
  It is well recognized that the peak wavelength
  of the 9.7$\mum$ extinction profile of
  silicate dust depends on the level of
  SiO$_4$ polymerization, e.g., it shifts
  from $\simali$9$\mum$ for pure SiO$_2$
  to $\simali$9.7$\mum$ for MgSiO$_3$ to
  $\simali$10.25$\mum$ for
  Mg$_{2.4}$SiO$_{4.4}$
  (J\"ager et al.\ 2003, Henning 2010).
  }

Various efforts have been made to observationally
determine the 9.7$\mum$ interstellar extinction feature.
However, the results were somewhat contradictory.
Kemper et al.\ (2004) analyzed
the $\simali$8--13$\mum$ spectrum
of the Galactic center source Sgr A$^{\ast}$
obtained with
the {\it Short Wavelength Spectrometer} (SWS)
on board the {\it Infrared Space Observatory} (ISO).
The 9.7$\mum$ silicate absorption profile
of Sgr A$^{\ast}$,
derived by subtracting a fourth-order polynomial
continuum from the observed spectrum,
peaks at $\simali$9.8$\mum$ and has
a FWHM of $\simali$1.73$\mum$,
much narrower than that of
``astronomical silicates". They suggested
that the narrow profile could be caused
by the contamination of the silicate emission
 intrinsic to the Sgr A$^{\ast}$ region.
Chiar \& Tielens (2006) obtained
the $\simali$2.38--40$\mum$ {\it ISO}/SWS spectra
of the diffuse ISM along the lines of sight toward
four heavily extinguished WC-type Wolf-Rayet stars.
They found that the 9.7$\mum$ silicate absorption
features of those four sources all peak at
$\simali$9.8$\mum$, but their widths vary
from $\simali$2.35$\mum$ for WR\,98a
to $\simali$2.7$\mum$ for WR\,104.

McClure (2009) derived the $\simali$5--20$\mum$
extinction curves for 28 G0--M4\,III stars lying behind
the Taurus, Chameleon I, Serpens, Barnard 59,
Barnard 68 and IC\,5146 molecular clouds.
This was achieved by comparing the spectrum
of each object obtained with the {\it Infrared Spectrograph} (IRS)
on board the {\it Spitzer Space Telescope}
with the stellar photospheric model spectrum
of Castelli et al.\ (1997).
She found that the 9.7$\mum$ silicate extinction feature
appears to be broadened in more heavily extinguished regions:
on average, in regions with $A_{\Ks}<1\magni$,\footnote{%
  McClure (2009) assumed $R_V=5$ which corresponds
  to $A_V\approx 7.75\,A_{\Ks}$.
  }
it peaks at $\simali$9.63$\mum$ and has a FWHM
of $\simali$2.15$\mum$; in contrast,  in regions
with $1<A_{\Ks}<7\magni$, it peaks at $\simali$9.82$\mum$
and has a FWHM of $\simali$2.72$\mum$.
Olofsson \& Olofsson (2011) derived the mid-IR
extinction curve for a highly obscured M giant
(\# 947) behind the dark globule B\,335
($R_V\approx4.9$, $\AV \sim10$\,mag),
using the $\simali$7--14$\mum$ {\it Spitzer}/IRS
spectrum complemented by the $\simali$5--16$\mum$
spectrum obtained with the ISOCAM/CVF instrument
on board {\it ISO}.
They found that the 9.7$\mum$ silicate extinction
feature peaks at $\simali$9.2$\mum$
and has a FWHM of $\simali$3.80$\mum$.
Also based on the {\it Spitzer}/IRS data,
van Breemen et al.\ (2011) investigated
the silicate absorption spectra of
three sightlines toward diffuse clouds
%(StRS 136, StRS 164, and StRS 354)
and four sightlines toward the Serpens,
Taurus and $\rho$ Ophiuchi molecular clouds.
They found that the 9.7$\mum$ silicate absorption
bands in the diffuse sightlines show a strikingly
similar band shape and all closely resemble
that of Sgr A$^\ast$ (Kemper et al.\ 2004),
while the 9.7$\mum$ features in the molecular sightlines
differ considerably from that of Sgr A$^\ast$
by peaking at $\simali$9.72$\mum$
and having a FWHM of $\simali$2.4$\mum$.
More recently, Fogerty et al.\ (2016)
analyzed the {\it Spitzer}/IRS spectra of
the 9.7$\mum$ silicate optical depths
of the diffuse ISM along the lines of sight
toward Cyg OB2 No.\,12, a heavily extinguished
luminous B5 hypergiant with $\AV\approx10.2\magni$,
and toward $\zeta$ Ophiuchi, a lightly extinguished
bright O9.5 star with $\AV\approx1\magni$.
They found appreciable differences between
the spectral profile of the 9.7$\mum$ silicate
absorption of Cyg OB2 No.\,12
and that of $\zeta$ Ophiuchi:
while the former peaks at $\simali$9.74$\mum$
and has a FWHM of $\simali$2.28$\mum$,
the latter peaks at $\simali$9.64$\mum$
and has a FWHM of $\simali$2.34$\mum$.
Moreover, the contrast between the feature
and the absorption continuum of the former
exceeds that of the latter by $\simali$30\%.

In this work, we shall make use of
the {\it Spitzer}/IRS spectra of
early-type O and B stars to derive
the 9.7$\mum$ interstellar silicate
extinction feature.
Differing from previous works,
we will utilize the ``pair'' method,
i.e., we will compare the {\it Spitzer}/IRS
spectra of two stars of similar spectral types,
with one star being obscured by interstellar dust
while the other star unaffected by dust.
In \S2 we present the sample selection
and in \S3 we describe the extinction
determination method.
The results are presented and discussed
in \S4 and summarized in \S5.

%\vspace{-5mm}
%
\section{The Sample}\label{sec:sample}
Typically, red giants are used to trace
the interstellar extinction in the IR
(e.g., see Indebetouw et al.\ 2005,
Flaherty et al.\ 2007, Gao et al.\ 2009,
Zasowski et al.\ 2009, Majewski et al.\ 2011,
Wang et al.\ 2013, Yuan et al.\ 2013,
Davenport et al.\ 2014, Xue et al.\ 2016).
This is because red giants are so bright
that they are still ``visible'' in the IR
even if they are severely obscured
in the UV/optical.
However, red giants are not suitable for
studying the 9.7$\mum$ interstellar silicate
extinction feature since the gas-phase molecular
absorption bands (e.g., the Si--O stretches of
the SiO gas which extend from $\simali$8$\mum$
to $\simali$11$\mum$) intrinsic to red giants
could contaminate the derivation of
the interstellar silicate feature
(e.g., see van Breemen et al.\ 2011).
In contrast, early-type stars are usually free
of circumstellar dust and are bright in the IR.
Also, their stellar photospheric emission in the IR
can be closely characterized by the Rayleigh-Jeans law.
Therefore, the determination of the stellar emission
in the IR of early-type stars is relatively straightforward
and accurate. In this work, we shall choose early-type
(O and B) stars as the tracers of
the interstellar silicate extinction profiles.

To date, $\simali$17,000 low-resolution spectra
have been obtained by {\it Spitzer}/IRS.
These spectra were merged from four slits:
SL2 ($\simali$5.21--7.56$\mum$),
SL1 ($\simali$7.57--14.28$\mum$),
LL2 ($\simali$14.29--20.66$\mum$), and
LL1 ($\simali$20.67--38.00$\mum$).
Chen et al.\ (2016) cross-identified
the types of these objects
in the {\it SIMBAD} database,
supplemented with the photometric results
from the {\it 2MASS} and {\it WISE} all-sky surveys.
This cross-identification resulted in
a database of 71 O stars,
271 B stars and 374 A stars.
We take the following approach
to select our sample:
\begin{itemize}
\item We first exclude those showing
      silicate and/or polycyclic aromatic
      hydrocarbon (PAH) emission features.
      Although early-type stars are generally
      free of circumstellar dust, some B- and
      A-type stars could host a protoplanetary
      or debris disk which could emit at
      the 9.7$\mum$ silicate feature
      (and even crystalline silicate
       features at 11.3, 18, 23, 28, 33,
       40 and 60$\mum$; see Henning \& Meeus 2011)
      and/or the 3.3, 6.2, 7.7, 8.6,
      and 11.3$\mum$ PAH features
      (see Seok \& Li 2017).
      This procedure leads to the rejection
      of 38 stars which show amorphous and/or
      crystalline silicate features
      and seven stars showing PAH features.
\item We require the signal-to-noise ratio (S/N)
      of the {\it Spitzer}/IRS spectrum to exceed 30.
      Because of this requirement,
      we are left with 48 O stars,
      214 B stars and 316 A stars.
\item To avoid saturation,
      for a star to be selected, we require
      it to be fainter than 5\,mag
      in the {\it 2MASS} $\Ks$ band,
      which is taken as the reference band
      for measuring the color excess.
      Because of this, Cyg OB2 No.\,12
      (an O star with a nominal {\it 2MASS}
      $\Ks$-band magnitude of $\simali$2.70$\magni$)
      and $\zeta$ Ophiuchi (a B star with a nominal
      $\Ks$ magnitude of $\simali$2.62$\magni$),
      which were used to trace the interstellar silicate
      extinction in Fogerty et al.\ (2016),
      are not included in our sample
      since they are much brighter than
      the saturation limit of $\simali$5$\magni$
      in $\Ks$.
\item To make sure that the line of sight
      is obscured sufficiently so that
      the 9.7$\mum$ silicate extinction
      is appreciable, we require the color index
      $J-\Ks$ to exceed 0.4$\magni$,
      corresponding to a line-of-sight
      extinction of $\AKs > 0.2\magni$,
      which is comparable to three times
      of the photometric error.

      While the peak extinction of
      the 9.7$\mum$ silicate feature is
      comparable to $\AKs$, the extinction of
      the blue and red wings of the 9.7$\mum$
      silicate feature drops by a factor of $>$2
      (e.g., see Xue et al.\ 2016).
      With $\AKs > 0.2\magni$, the entire 9.7$\mum$
      silicate extinction profile should be measurable.
\end{itemize}

The above selection process leads to
the selection of some A stars.
However, it turns out that those stars
have systematically small color excess
$E(J-\Ks)$ and weak silicate absorption.
We therefore decide to drop all those A stars.
Also, the free-free emission of supergiants,
arising from stellar winds of hot plasma,
may potentially deform
the slope of the stellar emission in the IR
and thus affect the derivation of
the interstellar extinction curve.\footnote{%
  Barlow \& Cohen (1977) reported
  the detection of IR excess at wavelengths
  longward of $\simali$10$\mum$
  in O supergiants and attributed it
  to free-free emission.
  Also, to account for the free-free emission
  from the stellar wind of the B5 hypergiant
  Cyg OB2 No.\,12, Fogerty et al.\ (2016)
  added an $\lambda^{-0.6}$ component
  in the IR spectrum.
  }
In the wavelength range of \emph{Spitzer}/IRS,
the ionized wind reveals itself through
the 7.5$\mum$ Pfund $\alpha$ and
12.4$\mum$ Humphries $\alpha$
emission lines of hydrogen.
Therefore, we exclude these stars
which exhibit the 7.5$\mum$ Pf$\alpha$
and 12.4$\mum$ Hu$\alpha$ emission lines.
As a result, the final sample consists
of five stars (i.e., one O star and four B stars).
Their locations in the Galaxy, spectral types,
brightness in the {\it 2MASS} and {\it WISE} bands
are listed in Table~\ref{table:t1}.
Also tabulated in Table \ref{table:t1}
is $Err(J-\Ks)$, the uncertainty of the {\it 2MASS} $J-\Ks$
color index, which is at most $\simali$0.03mag
for all the sightlines.
The \emph{Spitzer}/IRS spectra of
the selected stars are taken from
the Cornell Atlas of \emph{Spitzer}/IRS Sources
(CASSIS; Lebouteiller et al.\ 2011).
The CASSIS atlas provides
\emph{Spitzer}/IRS spectra
reduced using a dedicated spectral extraction pipeline
which performs optimal extraction for point-like sources
and a regular extraction for partially extended sources.
A comparison with the spectra reduced
from the \emph{Spitzer}/IRS pipeline
shows that the CASSIS spectra are better calibrated
(e.g., the negative flux which frequently shows up
in the {\it Spitzer}/IRS pipeline spectra
essentially never shows up in the {\it CASSIS} spectra).

\begin{figure*}
\vspace{-1mm}
\centering
\includegraphics[width={17.2cm}]{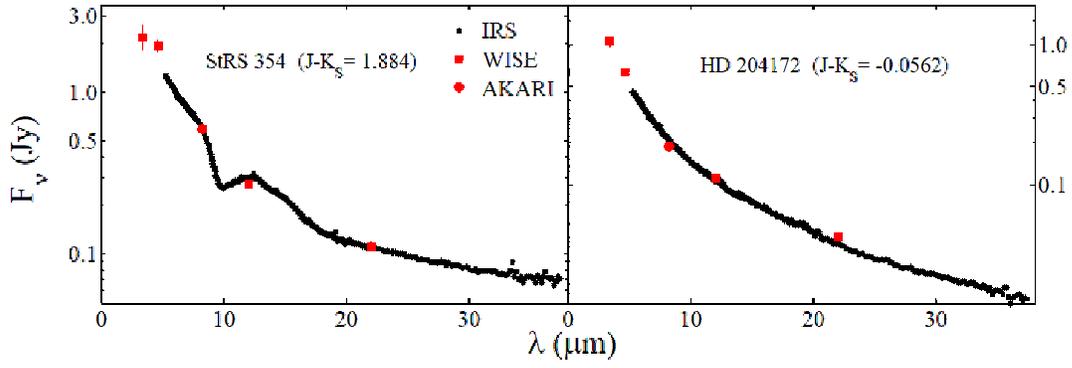}
%\vspace{20mm}

\caption{
\label{fig:otype}
The \emph{Spitzer}/IRS spectra and the IR photometry of the target O star StRS 354 and reference star HD 214680.
}
\vspace{4mm}
\end{figure*}

\begin{figure*}
\vspace{-1mm}
\centering
\includegraphics[width={18cm}]{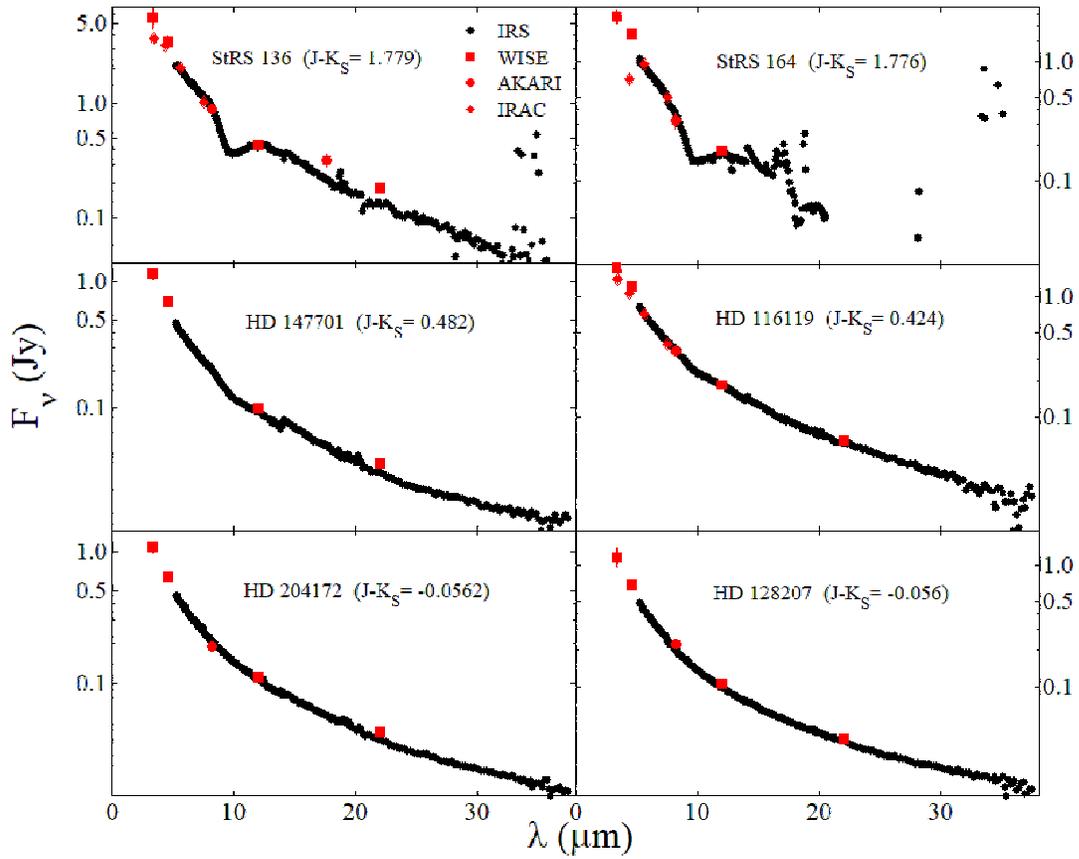}
\caption{ \label{fig:btype}
The \emph{Spitzer}/IRS spectra and the IR photometry of the  four target B stars
(StRS 136, StRS 164, HD 147701 and HD 116119) and the two reference stars (HD 204172 and HD 128207).  }
\vspace{-5mm}
\end{figure*}

Since the flux calibration of the {\it CASSIS}
spectra largely affects the reliability
of the derived silicate extinction feature,
we also use the IR photometric data
from {\it WISE}, {\it Spitzer}/IRAC
and {\it AKARI} to check the credibility of
the {\it CASSIS} spectra.
For each of our five target stars,
we search for the photometric data
in the four \emph{WISE} bands
at 3.35, 4.60, 11.56 and 22.08$\mum$ for
W1, W2, W3 and W4, respectively;
the four \emph{Spitzer}/IRAC bands
at 3.55, 4.49, 5.73 and 7.87$\mum$
for IRAC1, IRAC2, IRAC3, and IRAC4, respectively;
and the two \emph{AKARI} bands
at 8.23 and 17.61$\mum$ for
S9W and L18W, respectively.
The photometric data of the selected sources
are listed in Table~\ref{table:t2}.
As shown in Figures~\ref{fig:otype},\,\ref{fig:btype},
the photometry and spectroscopy of
of our five target stars as well as two reference stars
(see \S3) are highly consistent.
This also indirectly manifests that there is
no light variation for these sources
since the photometry and spectroscopy
were not performed at the same epoch.
The only exception is the W4-band
photometry of StRS\,164 which only
yields an upper limit. Also, the
{\it Spitzer}/IRS flux of StRS\,164
is negative at $\lambda > 20.5\mum$.
This implies that the measurement
at $\lambda\simgt20.5\mum$ may
not be reliable for StRS\,164.
In Table~\ref{table:t2}, we label
the W4-band flux of StRS\,164
with ``:'', which means ``uncertain''.
Nevertheless, all the sources show
high consistency between photometry
and spectroscopy at $\lambda < 15\mum$
(i.e., the wavelength range of
the {\it Spitzer}/IRS SL1 and SL2).
%This is not unexpected due to the increasing
%difficulty in the measurement
%toward longer wavelength.
As a result, the determination of
the 9.7$\mum$ silicate extinction profile
is apparently more reliable than
the 18$\mum$ profile.
In this work, we shall focus on
the extinction at $\lambda < 15\mum$.

\section{Method\label{sec:method}}
The interstellar UV/optical extinction curve
is often determined by comparing the spectrum
of a reddened star with that of an un-reddened star
of the same spectral type. We take a similar approach
for the 9.7$\mum$ silicate extinction.

The apparent stellar spectrum $F_\lambda$
is the intrinsic spectrum $F^0_{\lambda}$
dimmed by the interstellar extinction $A_\lambda$
and the geometrical distance $d$:
\begin{equation}
F_\lambda = \frac{F^0_{\lambda}\cdot
            \exp\left(-A_\lambda/1.086\right)
            \cdot \pi r^2}
            {4\pi d^2}~~,
\end{equation}
where $r$ is stellar radius.
If the reference star has
the same intrinsic spectrum
as the target star,
then the intrinsic flux ratio
of the star at $\Ks$ and $\lambda$
can be substituted by that of
the reference star, i.e.,
$F^0_\Ks/F^0_{\lambda} =
F^{\rm ref}_{\Ks}/F^{\rm ref}_\lambda$,
where ``ref'' refers to the reference star.
By comparing the observed spectrum of
the source star with the apparent spectrum
of the reference star,
we obtain $E(\lambda-\Ks)$,
the color excess between $\lambda$
and the $\Ks$ band, as
\begin{equation}\label{eq:EKslambda}
 \begin{aligned}
E(\lambda-\Ks) =  A_\lambda-A_\Ks
= -2.5\log\left(\frac{F_\lambda}{F_\Ks} \frac{F^{\rm ref}_\Ks}{F^{\rm ref}_\lambda}\right)\\
= -2.5\log\left(\frac{F_\lambda}{F^{\rm ref}_{\lambda}}\right)
+ 2.5\log\left(\frac{F_\Ks}{F^{\rm ref}_{\Ks}}\right) \\
= -2.5\log\left(\frac{F_\lambda}{F^{\rm ref}_{\lambda}}\right)
+ m^{\rm ref}_\Ks - m_\Ks ~~,
  \end{aligned}
\end{equation}
where $m_\Ks$ and $m^{\rm ref}_\Ks$
are respectively the apparent $\Ks$-band
magnitudes of the source star and the reference star.
Since we are considering the color excess
$E(\lambda-\Ks)$
instead of the extinction $A_\lambda$,
we do not need to know the distance $d$
to the source star (or the reference star)
as it gets cancelled out in eq.\,\ref{eq:EKslambda}.

\begin{figure*}
\vspace{1mm}
\centering
\includegraphics[width=.68\textwidth]{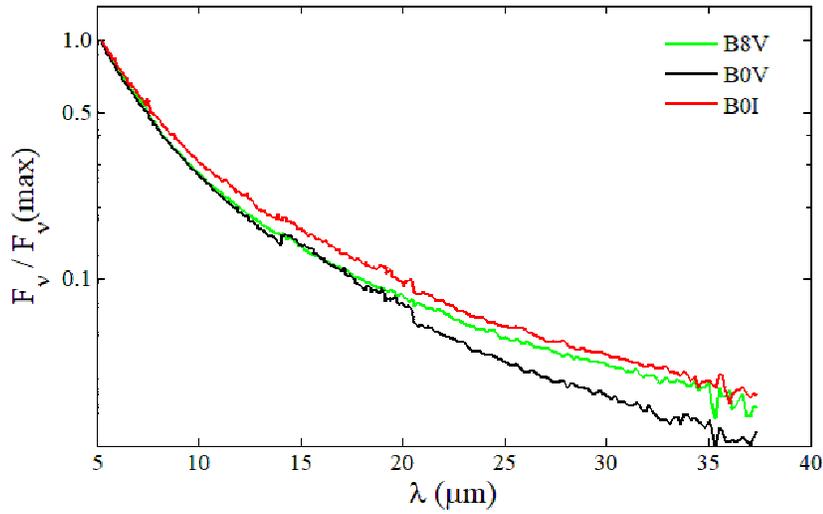}
\caption{\label{fig:refbijiao}
%Different influences of spectral type and luminosity class on the IRS spectrum of early-type stars.}
Comparison of the effects of the stellar spectral type on the IR slope of the stellar emission with that of the stellar luminosity class.}
\vspace{-5mm}
\end{figure*}

For each source star, we need to
find a reference star which is not
subject to any interstellar extinction
but has the same intrinsic spectrum
in the \emph{Spitzer}/IRS wavelength range.
In principle, an unreddened star
of the same spectral type
{\it and} the same luminosity class
would be the best choice.
For this purpose, we first attempt to
select unreddened O and B stars
by requiring the observed color index
$J-\Ks$ to be smaller than 0,
i.e., bluer than an A0 star.\footnote{%
  By definition, for A0 stars $J-\Ks = 0$.
  }
However, this turns out to be impractical
as it is difficult to find a reference star
of the same spectral type
and the same luminosity
class for every source star.
Specifically, all the source stars
selected in this work
are supergiants except HD\,147701
which is classified as a giant,
while most of the stars
with $J-\Ks < 0$ are dwarfs.
Therefore, it is desirable to somewhat
relax the criteria: perhaps a reference
star could be selected based on its
spectral type {\it or} luminosity class alone?
To this end, we explore which plays
a more important role,
the luminosity class or the spectral type,
in affecting the slope of
the stellar continuum in the IR,
by comparing the stellar emission
spectral energy distribution
in the wavelength range of {\it Spitzer}/IRS
for stars of the same luminosity class
or of the same spectral type.
As illustrated in Figure~\ref{fig:refbijiao},
the spectral slope of a B0V star
in the wavelength range of $\simali$5--38$\mum$
closely resembles that of a B8V star.
In contrast, the spectral slope of a B0V star
considerably differs from that of a B0I star.
This demonstrates that the luminosity class
affects the IR slope of the stellar emission
more significantly than the spectral type.\footnote{%
  This actually can be understood
  in terms of the Rayleigh-Jeans approximation,
  i.e., the stellar spectral type or effective
  temperature ($\Teff$) has little effect
  on the IR slope of the stellar emission:
  $F_\lambda \propto \lambda^{-4}\,T$
  provided $hc/\left(\lambda k \Teff\right) \ll 1$,
  where $h$ is the Planck constant,
  $c$ is the speed of light,
  and $k$ is the Boltzman constant.
  }
Therefore, we choose HD\,204172,
a supergiant with $J-\Ks\approx-0.056$
and $\alpha\approx1.768$
(see Table~\ref{table:t1}),
as the reference star for
the four supergiants
(i.e., one O supergiant
and three B supergiants)
among our five source stars,\footnote{%
  Koornneef (1983) measured
  the intrinsic color of B0I stars
  to be $J-K\approx-0.11$.
  This suggests that HD\,204172
  may suffer a reddening of
  $E(J-\Ks) \approx 0.05\magni$.
  Consequently, this may bring up
  a reddening of $E(\lambda-\Ks)\approx0.01\magni$
  in the {\it Spitzer}/IRS wavelength range,
  which is negligible.
  }
where $\alpha\equiv d\ln F_{\nu}^{\star}/d\ln\nu$
is the slope of the $\simali$5--38$\mum$
stellar emission, where $F_\nu^{\star}$ is
the stellar flux at frequency $\nu$.
HD\,147701, a B5III star, is the only target star
in our sample which is not classified as a supergiant.
We could not find any reference star with the same
luminosity class as HD\,147701.\footnote{%
   We note that the classification of HD\,147701
   as a giant may be questionable since B5 stars
   are usually dwarfs or supergiants.
   }
Therefore, we take HD\,128207,
a B8V star with $J-\Ks\approx-0.056$\footnote{%
  This color index agrees rather well with
  that of Koornneef (1983) who measured
  $J-K\approx-0.04$ for B8V stars.
  %The agreement is even more promising
  %if one accounts for the difference
  %between the $K$ and $\Ks$ filters.
  }
to be the reference star for HD\,147701.
Finally, we note that the integrated CO\,(1--0)
line intensities along the lines of sight
to HD\,204172 ($\simali$0.03$\K\km\s^{-1}$)
and HD\,128207 ($\simali$0.04$\K\km\s^{-1}$)
are rather small, as measured by {\it Planck}
(see Planck Collaboration XIII 2014
and Table~\ref{table:t1}).
This confirms that these reference stars
experience no or negligible extinction.

\begin{figure*}
\vspace{3mm}
\centering
\includegraphics[width=.68\textwidth]{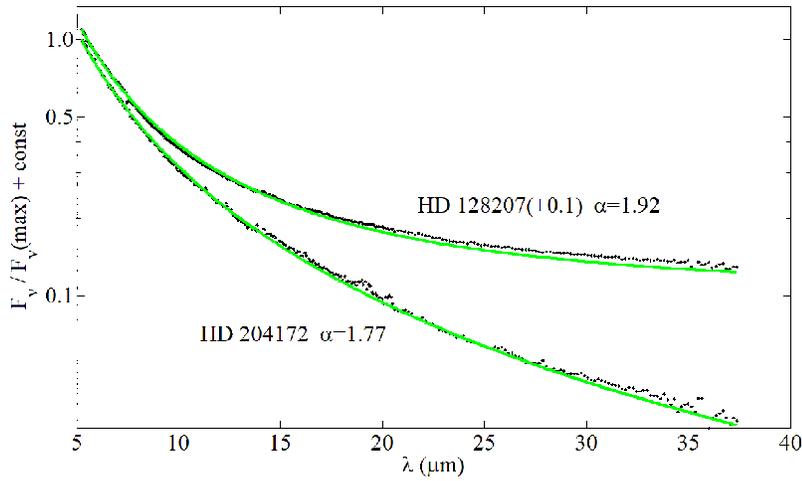}
\caption{\label{fig:refduibi}
%The normalized spectrum (black dots) of the reference stars with a shift for clarification and the power law fitting (green line) with the power law index. }
The \emph{Spitzer}/IRS spectra (black dots) of the two reference stars (HD 128207 and HD 204172)
and their power-law representation (i.e., $F_\nu \propto \nu^{\alpha}$)}
\vspace{-5mm}
\end{figure*}

\begin{figure*}
\vspace{-2mm}
\centering
\includegraphics[width=.68\textwidth]{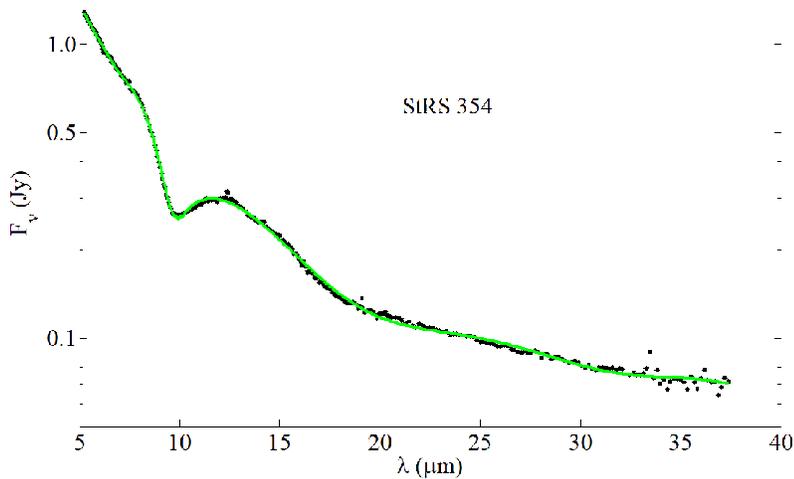}
\caption{ \label{fig:sourcenihe}
%The example of the spectrum (black dots) of one target star and the polynomial fitting (green line) of the continuum. }
The \emph{Spitzer}/IRS spectrum (black dots) of StRS 354, a target star, and its polynomial fit (green line). }
\vspace{-5mm}
\end{figure*}

There are some small-amplitude spikes
in the {\it Spitzer}/IRS spectra of
both reference and source stars due to noise,
although some may be ionic or atomic lines.
These small spikes cause appreciable fluctuations
in the color excess because the color excess
comes from the division of the two spectra
where the error propagation amplifies the uncertainty.
Since we are mostly interested in the silicate feature,
we smooth the observed spectra in the following way:
(i) for the reference stars, the spectra
are fitted with a power law
(see Figure~\ref{fig:refduibi});
(ii) for the source stars, the spectra
are fitted with a polynomial function
(e.g., see Figure~\ref{fig:sourcenihe}).
In this way, we may lose the information
of some dust species (e.g., ices),
but the profile of the silicate extinction
will be better determined.

\section{Results and Discussion}\label{sec:results}
By comparing the {\it Spitzer}/IRS spectra of
our five source stars with that of reference stars,
we calculate the color excess $E(\lambda - \Ks)$.
We further normalize $E(\lambda - \Ks)$
by $E(J-\Ks)$ to cancel out the extinction quantity.
In Figure~\ref{fig:extinccon} we show
the color excess ratios $E(\lambda - \Ks)/E(J-\Ks)$
for our five sources.
Most noticeably, the 9.7$\mum$ silicate extinction
feature is pronounced in all five sources.

\subsection{The 9.7$\mum$ Silicate Extinction Profile}

In Figure~\ref{fig:SilNorm}
we compare the extinction profiles of
our five sources which are all normalized
to their peak extinction at $\simali$9.7$\mum$.
It is apparent that the 9.7$\mum$ silicate profiles
exhibit appreciable variations in band width,
with that of HD\,147701 and StRS\,164 being
considerably wider than that of
StRS\,136 and StRS\,354.
%HD\,116119 in between (1.90um).
%

\begin{figure*}
\vspace{7mm}
\centering
\includegraphics[width=0.65\textwidth]{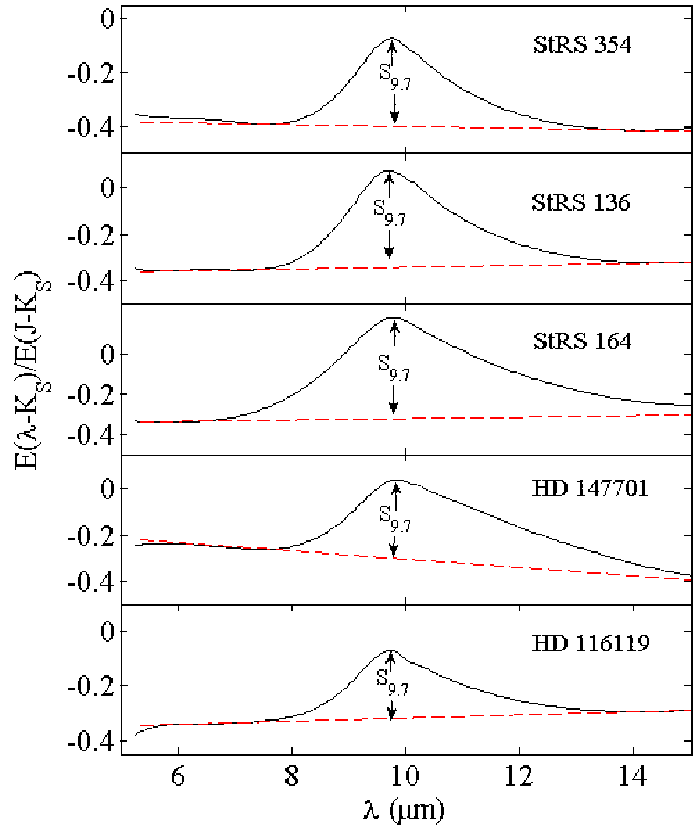}
\caption{\label{fig:extinccon}
%The silicate extinction profile with the linear continuum that is subtracted (dash line).}
The 9.7$\mum$ silicate extinction profiles of our five target stars. Also shown is the underlying continuum extinction (dashed line).    }
\vspace{-5mm}
\end{figure*}

\begin{figure*}
\vspace{7mm}
\centering
\includegraphics[width=0.68\textwidth]{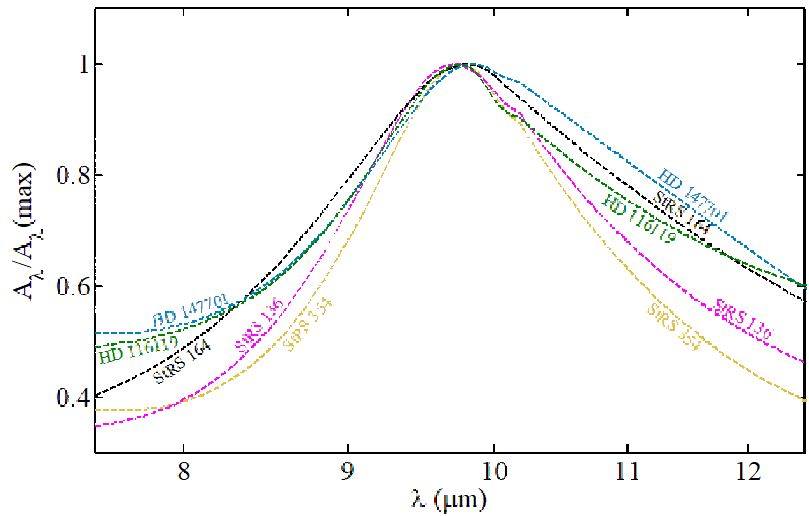}
\caption{\label{fig:SilNorm}
              Comparison of the 9.7$\mum$
              silicate extinction profiles of
              our five target stars shown in
              Figure~\ref{fig:extinccon}.
              All profiles are normalized
              to their peak extinction at
              $\simali$9.7$\mum$.
              %Note that here the continuum
              %extinction defined in Figure~\ref{fig:extinccon}
              %is not subtracted.
              }
\vspace{-5mm}
\end{figure*}

%
%
To be more quantitative,
we determine the peak wavelength ($\lambdapeak$)
and FWHM ($\gammasil$) of the 9.7$\mum$ silicate
extinction feature for each source by fitting
the $E(\lambda - \Ks)/E(J-\Ks)$ color excess curve
with a Gaussian function peaking around 9.7$\mum$
combined with a linear function representing
the continuum extinction
underlying the silicate feature
(see Figure~\ref{fig:extinccon}).

We tabulate in Table~\ref{table:FWHM}
the fitted $\lambdapeak$
and $\gammasil$ parameters,
and  strengths ($S_{9.7}$)
of the 9.7$\mum$ silicate extinction profiles
feature derived from Figure~$\ref{fig:extinccon}$.

It is apparent that the FWHM of
the 9.7$\mum$ extinction feature
varies among our five sources.
It appears that they fall into two groups:
$\gammasil\approx2.0\mum$ for
StRS\,136, StRS\,354 and HD\,116119,\footnote{%
    The red wing of the 9.7$\mum$ feature
    of HD\,116119 is rather flat
    (see Figure~$\ref{fig:extinccon}$)
    and the continuum extinction underneath
    the 9.7$\mum$ feature could have been
    overestimated and thus the derived
    band width could have been somewhat
    underestimated (see Figure~\ref{fig:SilNorm}).
    }
and $\gammasil\approx3.0\mum$ for
StRS\,164 and HD\,147701.
%\footnote{\textbf{StRS\,136, StRS\,354 and StRS\,164,
%this three sources are also researched by
%van Breemen et al. (2011).
%They divide the total extinction into
%a ¡°continuum¡± component and the silicate
%absorption features. Assuming the
%intrinsic spectrum of the background
%star to be approximately a Rayleigh-Jeans tail,
%they defined the intrinsic continuum for
%the 9.7$\mum$ silicate feature
%by the flux levels at 7.5$\mum$ and 12.3$\mum$,
%fit it by a power-law with wavelength
%as well. However we choose the reference stars
%without extinction as the intrinsic spectra.
%As a result, they
%concluded that the silicate absorption
%profiles at 9.7$\mum$ were similar with
%the feature observed towards the Galactic
%centre (Kemper et al. 2004).}}
This is consistent with the earlier findings
of Roche \& Aitken (1984, 1985)
who found that, with $\gammasil\approx2.4\mum$,
the silicate extinction profiles along the lines of sight
to diffuse clouds and to the Galactic center are narrow,
resembling the circumstellar silicate emission profile
of the supergiant $\mu$ Cep,
while the silicate extinction profiles
of dense molecular clouds,
with $\gammasil\approx3.4\mum$,
are significantly broader,
similar to the Trapezium emission profile.
However, different results have also been
reported in the literature.
Rieke \& Lebofsky (1985) derived
$\gammasil\approx3.2\mum$
for the 9.7$\mum$ silicate extinction profile
of the Galactic center sightlines
toward three stars within
$\simali$100$^{\prime\prime}$
($\simali$5\,pc) of
the Galactic center,\footnote{%
   Rieke \& Lebofsky (1985)
   computed the interstellar extinction law
   between 8 and 13$\mum$ by assuming
   the $\simali$8--13$\mum$ interstellar
   opacity profile to resemble the average
   of the emission spectra of $\mu$ Cep
   (Russell et al.\ 1975) and
   the Trapezium (Forrest et al.\ 1975)
   and adopting $A_V/\Delta\tau_{9.7}=16.6$,
   where $\Delta\tau_{9.7}$ is the optical depth
   of the 9.7$\mum$ silicate absorption.
   }
which is much broader than
that of  Roche \& Aitken (1985),
obtained for several mid-IR sources
within $\simali$2\,pc of
the centre of the Galaxy.
While Pegourie \& Papoular (1985) determined
$\gammasil\approx2.2\mum$
for dense clouds, which is much narrower than
that of Roche \& Aitken (1984).
See Figure~3 of Draine (1989)
for a detailed comparison.

For our five sources, there does not seem
to be any relation between $\gammasil$
and the interstellar environment.
Both highly reddened lines of sight
for which $E(J-\Ks)\approx1.9\magni$
(i.e., StRS\,136, StRS\,354)
and a moderately obscured line of sight
with  $E(J-\Ks)<0.5\magni$
(i.e., HD\,116119)
exhibit a narrow 9.7$\mum$ silicate
extinction profile.
Similarly, a broad 9.7$\mum$ profile
is also seen in both highly extinguished
star (i.e., StRS\,164) and moderately
reddened star (i.e., HD\,147701).
McClure (2009) found that
the 9.7$\mum$ silicate extinction profile
broadens at larger extinction of $\AKs > 1.0$,
based on a comparison of the \emph{Spitzer}/IRS
spectra of obscured G0--M4 stars
with their stellar model atmospheric spectra.
We note that some of her stars (e.g., M stars)
may have intrinsic circumstellar silicate emission.
%Moreover, star-forming regions in her study have
%deep extinction out of our reach.
%
Some of the sightlines considered here
may trace dense clouds.
The three most reddened sources
(StRS\,136, StRS\,164 and StRS\,354) have
a color index of $J-\Ks\sim1.8\magni$
which corresponds to $\AV\sim10\magni$.
They possibly traverse some dense medium.
Indeed, the integrated CO (1-0) line intensities
from {\it Planck} observations (Planck Collaboration XIII
2014), $\simali$425$\K\km\s^{-1}$ and
$\simali$186$\K\km\s^{-1}$, respectively,
for the StRS\,136 and StRS\,164 sightlines,
are about one order of magnitude
higher than that of the other sightlines
(see Table \ref{table:t1}).
As a matter of fact, these two sources
have the largest color excess ratio at 9.7$\mum$,
i.e., $E(9.7\mum-\Ks)/E(J-\Ks) \approx0.076$
and 0.181, respectively.
%, and the corresponding extinction curves
%lie above the model curves for $\RV=$5.5.
On the other hand, the CO line intensity of
the StRS\,354 sightline is only
$\simali$18.9$\K\km\s^{-1}$,
one order of magnitude smaller than
that of StRS\,136 and StRS\,164.
%One possibility is that StRS\,354 is much
%more farther that the CO cloud.
%Considering that this star is an O-type star
%while the other two are B-type stars,
%such possibility exists since
%O star is much brighter.
With a much smaller color excess ratio,
i.e., $E(9.7\mum-\Ks)/E(J-\Ks)\approx-0.073$,
it is likely that StRS\,354 traces diffuse medium
and its large color excess may originate
from a pile of diffuse clouds along the sightline.
%
%There is another possibility that
%cannot be completely ruled out,
%i.e. StRS 354 has some circumstellar dust
%that contributes to the reddening.
%Our first step of selecting the tracers
%excluded those whose IRS spectrum exhibits
%the emission features either from silicate or
%from carbonaceous (mainly PAH) dust.
%StRS 354 does not show any dust feature.
%The circumstellar dust around early-type stars
%is usually optically thin,
%and the circumstellar silicate emission
%would be difficult to recognize
%when it is mixed with and reduces
%the interstellar extinction
%(but see discussion in Section 4.2).
%\citet{Breemen11} also studied the three sightlines.
%
The sightlines toward HD\,116119
and HD\,147701
unambiguously trace diffuse medium.
They exhibit a color index
of $J-\Ks\approx0.40$
and $J-\Ks\approx0.48$, respectively,
corresponding to $\AV\sim2-3\magni$
(see Table \ref{table:t1}).
They also show small integrated CO intensities
($\simali$6.0 and $\simali$13.5$\K\km\s^{-1}$ respectively).

Interestingly, the peak wavelengths of
the 9.7$\mum$ extinction feature of
our five source stars are rather stable:
$\lambdapeak\approx9.73\mum$ for
all the three sightlines
with $\gammasil\approx2.0\mum$, and
$\lambdapeak\approx9.84\mum$ for
the two sightlines with $\gammasil\approx3.0\mum$.
%The dispersion in $\lambdapeak$ is rather small.
%
Although it appears that $\lambdapeak$ intends
to slightly increase with $\gammasil$,
this cannot be explained in terms of
the grain size effect (e.g., for a broadening of
$\Delta\gammasil=1\mum$,
``astronomical silicate'' would shift the peak of
the 9.7$\mum$ profile by
$\Delta\lambdapeak\approx0.24\mum$;
see Shao et al.\ 2017).

\begin{figure*}
\vspace{5mm}
\centering
\includegraphics[width=.68\textwidth]{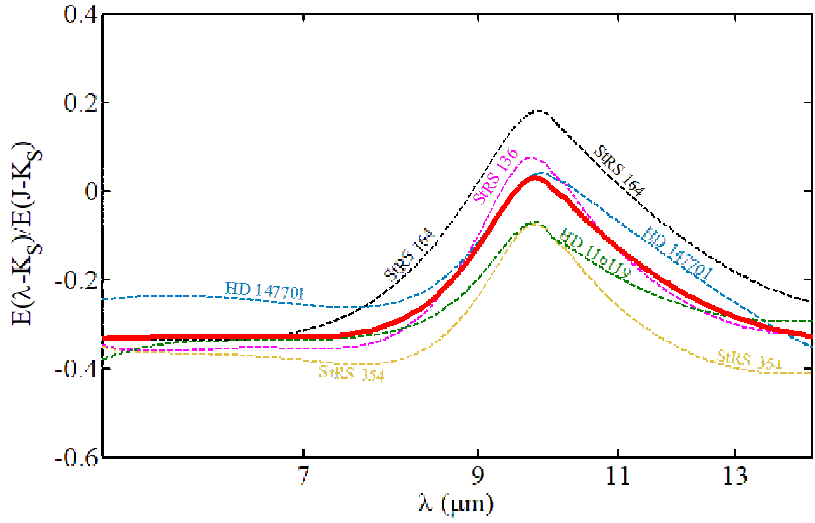}
\caption{ \label{fig:obme2}
The 9.7$\mum$ silicate
extinction profiles derived for the five target stars (broken line)
and their average (thick solid red line).}
\vspace{-5mm}
\end{figure*}

\begin{figure*}
\vspace{5mm}
\centering
\includegraphics[width=.68\textwidth]{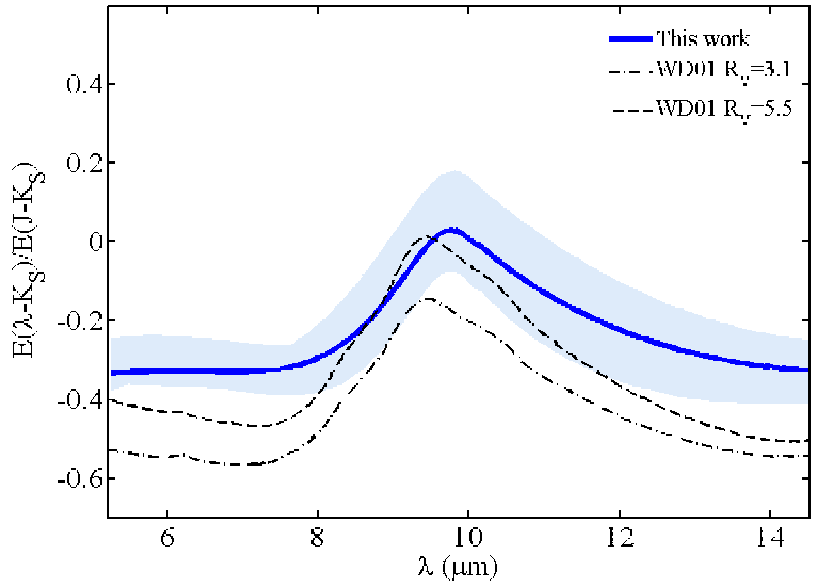}
\caption{ \label{fig:fobwd01m}
Comparison of the mean 9.7$\mum$ silicate extinction curve
derived in this work (thick blue) with that of the WD01 model
for $R_V=3.1$ (dotted line) and for $R_V=5.5$ (dashed line).
The shaded region represents the variation range
of the extinction derived here.
}
\vspace{-5mm}
\end{figure*}

\begin{figure*}
\vspace{5mm}
\centering
\includegraphics[width=0.68\textwidth]{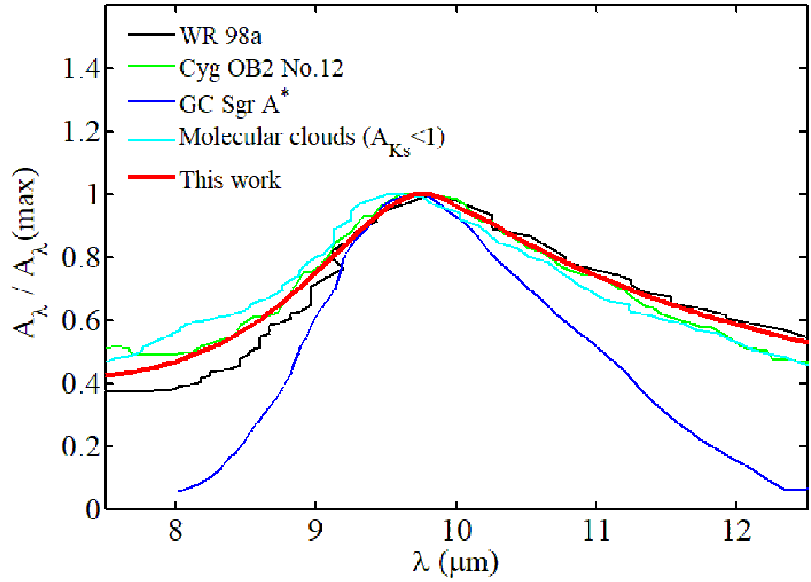}
\caption{ \label{fig:profilevs3}
Comparison of the mean 9.7$\mum$ silicate extinction profile derived for
the five early-type target stars (red line) with that of
(i) the diffuse ISM toward WR98a (black line; Chiar \& Tielens 2006),
(ii) the diffuse ISM toward Cyg OB2 No.\,12 (green line;
     Fogerty et al.\ 2016),
(iii) the Galactic Center sightline toward Sgr A$^\ast$
      (blue line; Kemper et al.\ 2004), and
(iv) molecular clouds of $\AKs<1\magni$
      (cyan line; McClure 2009). }
\vspace{-5mm}
\end{figure*}

\begin{figure*}
\vspace{1mm}
\centering
\includegraphics[width=0.69\textwidth]{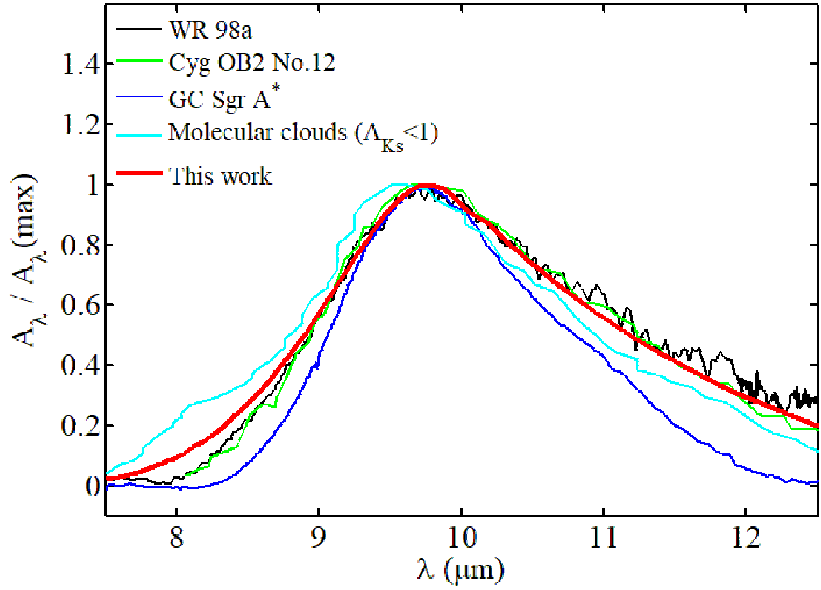}
\caption{\label{fig:profilevs}
               Same as Figure~\ref{fig:profilevs3}
               but with the continuum extinction
               underlying the 9.7$\mum$ silicate feature
               subtracted.
               }
\vspace{5mm}
\end{figure*}

In Figure~\ref{fig:obme2} we compare
the 9.7$\mum$
silicate extinction  profiles of all five sources.
It is apparent that the dispersion in $\lambdapeak$
is rather small. The mean profile, obtained by
averaging over all five sources, peaks at
$\lambdapeak\approx9.75\mum$
and has a width of $\gammasil\approx2.2\mum$.
In Figure~\ref{fig:fobwd01m} we compare
the mean 9.7$\mum$ extinction profile
with that of the WD01 models for $R_V=3.1$
and $R_V=5.5$. The 9.7$\mum$ profiles
of the WD01 model
(i.e., $\gammasil\approx3.0\mum$ and
$\lambdapeak\approx9.51\mum$ for $R_V=3.1$,
and  $\gammasil\approx3.6\mum$ and
$\lambdapeak\approx9.45\mum$ for $R_V=5.5$)
are much broader and peak at a much shorter wavelength
than the mean extinction profile derived in this work.
This is not unexpected since the dielectric functions
of ``astronomical silicate'' were synthesized
based on the Trapezium emission profile
which peaks at $\lambdapeak\approx9.56\mum$
and has a width of $\gammasil\approx3.45\mum$
(see Gillett et al.\ 1975, Draine \& Lee 1984).
Figure~\ref{fig:profilevs3} compares
the mean extinction profile of
the 9.7$\mum$ feature derived here
(normalized to its maximum value)
with that of Sgr A$^{\ast}$,
a Galactic center source (Kemper et al.\ 2004),
WR\,98a, a heavily extinguished
Wolf-Rayet star (Chiar \& Tielens 2006),
Cyg OB2 No.\,12, a heavily extinguished
hypergiant (Fogerty et al.\ 2016),
and those molecular clouds
of McClure (2009) with $\AKs<1\magni$.
Figure~\ref{fig:profilevs} is the
the same as Figure~\ref{fig:profilevs3},
but with the continuum extinction underlying
the 9.7$\mum$ feature subtracted.
It is apparent that the mean silicate profile
derived in this work is appreciably broader than
that of the GC source Sgr A$^{\ast}$,
no matter the underlying continuum
extinction is subtracted or not.
It is in close agreement with that of
WR\,98a and Cyg OB2 No.\,12,
both of which are typical diffuse lines of sight.
In contrast, the 9.7$\mum$ silicate profile
of McClure (2009) for molecular clouds
with $\AKs<1\magni$ peaks at
$\lambdapeak\approx9.63\mum$,
although its width ($\gammasil\approx2.15\mum$)
is close to that of WR\,98a, Cyg OB2 No.\,12,
and the one derived here.
The discrepancy in the silicate
absorption profiles may arise from
grain size, shape, and more importantly,
composition.

The extinction around the 9.7$\mum$
silicate feature, expressed as the color
excess ratio at $\lambdapeak$,
$E(\lambdapeak-\Ks)/E(J-\Ks)$,
varies from $\simali$$-0.1$
to $\simali$0.2, with a mean ratio of
$\langle E(\lambdapeak-\Ks)/E(J-\Ks)\rangle\approx0.03\pm0.11$
(see Table~\ref{table:t5}).
In comparison, the WD01 model predicts
$E(\lambdapeak-\Ks)/E(J-\Ks)\approx-0.162$
for $\RV=3.1$ and
$E(\lambdapeak-\Ks)/E(J-\Ks)\approx-0.008$
for $\RV=5.5$.
As shown in Figure~\ref{fig:fobwd01m},
the color excess ratio at $\lambdapeak$
derived here agrees better with
that of the WD $\RV=5.5$ model.
This is also true for the extinction
ratio $A_\lambda/\AKs$ around the 9.7$\mum$
feature derived by Xue et al.\ (2016)
from the photometric data of
\emph{Spitzer}/IRAC,
\emph{Spitzer}/MIPS and \emph{WISE}
in the sense that it better matches
the WD01 $\RV=5.5$ model
(see Figure~20 in Xue et al.\ 2016).
%On the other hand, the color excess
%ratio $E(\lambdapeak-\Ks)/E(J-\Ks)$
%or the extinction ratio $A_\lambda/\AKs$
%is systematically higher than that derived
%by McClure (2009).
%
We attribute the variations in
$E(\lambdapeak-\Ks)/E(J-\Ks)$
to grain size effects: for dense regions
of $\RV=5.5$, large grains of $\simgt$0.5$\mum$
will raise the continuum extinction underlying
the 9.7$\mum$ silicate feature and thus
increase $E(\lambdapeak-\Ks)/E(J-\Ks)$
and $A_\lambda/\AKs$.
Indeed, as mentioned earlier,
StRS\,136 and StRS\,164,
the two sightlines which
exhibit the largest
$E(\lambdapeak-\Ks)/E(J-\Ks)$
color excess ratios,
likely traverse dense clouds
as revealed by the high CO (1-0)
line intensities.
According to the newest catalog of
the Galactic molecular clouds
(Rice et al.\ 2016),
there are molecular clouds very
close to the sightlines
toward StRS\,136 and StRS\,164.
In addition, the variations among the fractional
contributions of other dust components
(e.g., carbon dust) to the continuum extinction
underneath the 9.7$\mum$ silicate feature
could also cause the variations of the color
excess ratios at $\lambdapeak$:
a higher fractional contribution of carbon dust
to the continuum extinction could raise
$E(\lambdapeak-\Ks)/E(J-\Ks)$.
%
%Van Breemen et al.\ (2011) argued that
%these sightlines as representative
%of diffuse environment (see their Table 1)
%based on the relativity to very dense molecular clouds
%(Serpens, Taurus and $\rho$ Ophiuchi).

%%%%%

Finally, we have also explored
$A_V/\Delta\tau_{9.7}$ for our five sources.
While it is relatively straightforward to
derive $\Delta\tau_{9.7}$, the 9.7$\mum$
silicate absorption optical depth,
from the observed, continuum-subtracted
9.7$\mum$ extinction profile, it is less
straightforward to determine $A_V$,
the visual extinction toward each of
the source sightline.
For each source star, we first calculate
the $V-\Ks$ color from its {\it 2MASS}
$\Ks$ magnitude and its {\it SIMBAD}
$V$ magnitude and then compare it
with the intrinsic $\left(V-\Ks\right)_0$ color
from the Allen's Astrophysical Quantities (Cox 2000)
for standard stars of the same spectral
and luminosity types.
This allows us to estimate the color excess
$E(V-\Ks)$.
If $\AKs$ is known, then one can derive
$A_V$ from $\AKs$ and $E(V-\Ks)$:
$A_V = \AKs + E(V-\Ks)$.
We determine $\AKs$
from the $E(J-\Ks)$ color excess
based on the {\it 2MASS} data
and adopt $A_J/A_\Ks = 2.72$
(Xue et al.\ 2017):
$\AKs\approx E(J-\Ks)/1.72$.
See Table~\ref{table:VKs}
for the color and color excess data
derived for our five source stars.
In Table~\ref{table:FWHM} we tabulate
$A_V$,  $\Delta\tau_{9.7}$,
and $A_V/\Delta\tau_{9.7}$
for each source.
It is apparent that,
with an average ratio of
$\langle \AV/\Delta\tau_{9.7}\rangle\approx18.2$,
they are in close agreement
with that of the solar neighborhood
diffuse ISM for which
$\AV/\Delta\tau_{9.7}\approx18.5$
(Roche \& Aitken 1984),
but substantially exceeding
that of the Galactic center
($\AV/\Delta\tau_{9.7}\approx9$,
Roche \& Aitken 1985) and
the dust torus around active
galactic nuclei
(AGN; $\AV/\Delta\tau_{9.7}\approx5.5$,
Lyu et al.\ 2015).
This suggests that the silicate grains
in the interstellar clouds toward
our five source stars could be considerably
smaller than that toward the Galactic center
and the AGN dust torus
(see Shao et al.\ 2017).

%%%%%

\subsection{Interstellar or Circumstellar?}
As described in \S\ref{sec:sample},
the target sources are selected by excluding
those showing the emission features of
circumstellar amorphous silicates,
crystalline silicates, and PAHs.
However, circumstellar amorphous carbon
or graphite could be present without being
detected since it is featureless in the IR
(e.g., see Rouleau \& Martin 1991,
Draine 2016).\footnote{%
  Pure crystalline graphite has a weak feature
  at $\simali$11.52$\mum$ arising from
  the out-of-plane lattice resonance of graphite
  (Draine 1984). This feature has not been detected
  yet, e.g., by {\it ISO}/SWS or {\it Spitzer}/IRS.
  It may be observable with the MIRI spectrograph
  on {\it James Webb Space Telescope}
  (see Draine 2016).
  }
Also, circumstellar silicate could be present
but its emission is hidden by the interstellar
silicate extinction.
To examine whether our sources could be
surrounded by circumstellar dust, we follow
van der Veen \& Habing (1988) who proposed
that the ratio of the {\it IRAS} 12$\mum$ flux
to the {\it IRAS} 25$\mum$ flux could be used
as an effective diagnosis of the presence
or absence of circumstellar dust.
Unfortunately, none of our five target sources
was detected in the {\it IRAS} 12, 25, 60 and
100$\mum$ bands.
We therefore, instead, use the fluxes of the {\it WISE}
W3 and W4 bands since they are closely
similar to the {\it IRAS} 12 and 25$\mum$ bands.
To this end, the criterion of van der Veen \& Habing (1988)
for the presence of circumstellar dust becomes
W3$-$W4\,$>$\,1.0
(Jarrett et al. 2011; Sjouwerman et al. 2009).\footnote{%
  For example, Cyg OB2 No.\,12,
  a prototypical diffuse sightline
  and known to be free of circumstellar dust,
  shows W3$-$W4\,$\approx$\,0.4.
  }
As can be seen in Table~\ref{table:t2},
all our targets show W3$-$W4\,$<$\,0.3
(except StRS\,164 for which the {\it WISE} W4 flux
may be uncertain), suggesting that our sources
are free of circumstellar dust.

\begin{figure*}
\vspace{-2mm}
\centering
\includegraphics[width=\textwidth]{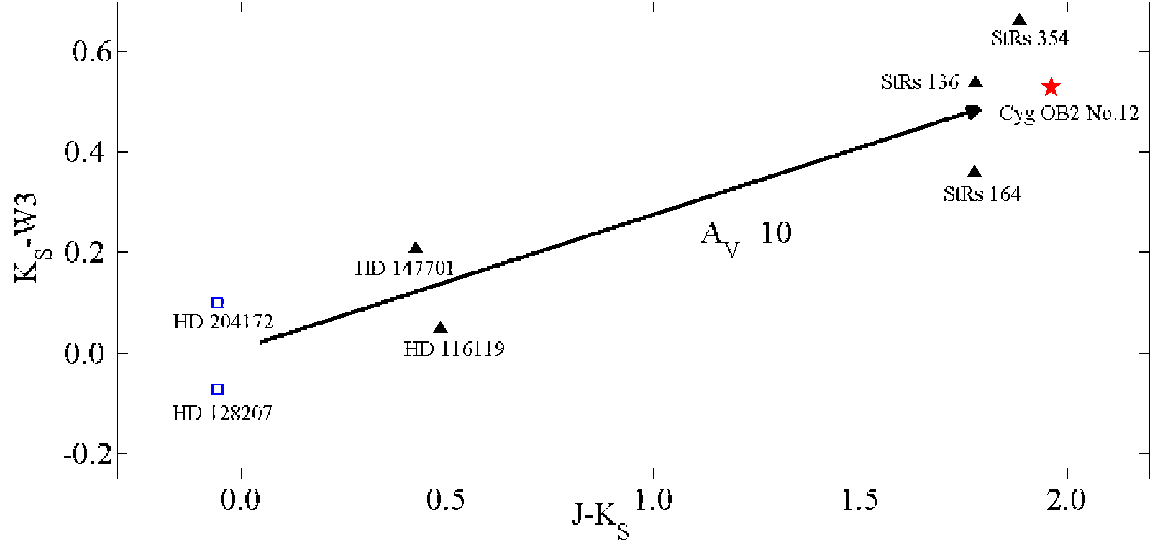}
\caption{\label{fig:fkw3jk}
The $\rm J-\Ks$ vs. $\rm \Ks-W3$ diagram for the five target stars (triangles),
two reference stars (open squares), and Cyg OB2 No.\,12 (red star).
The arrowed straight line denotes the extinction tendency for $\AV = 10$ mag (see Xue et al. 2016).}
\vspace{-5mm}
\end{figure*}

Xue et al.\ (2016) found that those stars
with circumstellar silicate emission appear
to display apparent color excess at $\Ks-{\rm W}3$
in the $J-\Ks$ vs. $\Ks-{\rm W}3$ diagram
(see Figure~19 in Xue et al.\ 2016).
We show in Figure~\ref{fig:fkw3jk}
the distribution of our five target stars
in the $J-\Ks$ vs. $\Ks-{\rm W}3$ diagram,
together with the average extinction tendency
derived by Xue et al.\ (2016).
It is noteworthy that all our stars
follow the average extinction law
reasonably well, indicating that our sources
are free of circumstellar dust.

\subsection{Analytical Representation of the Extinction}
\begin{figure*}
\vspace{5mm}
\centering
\includegraphics[width=\textwidth]{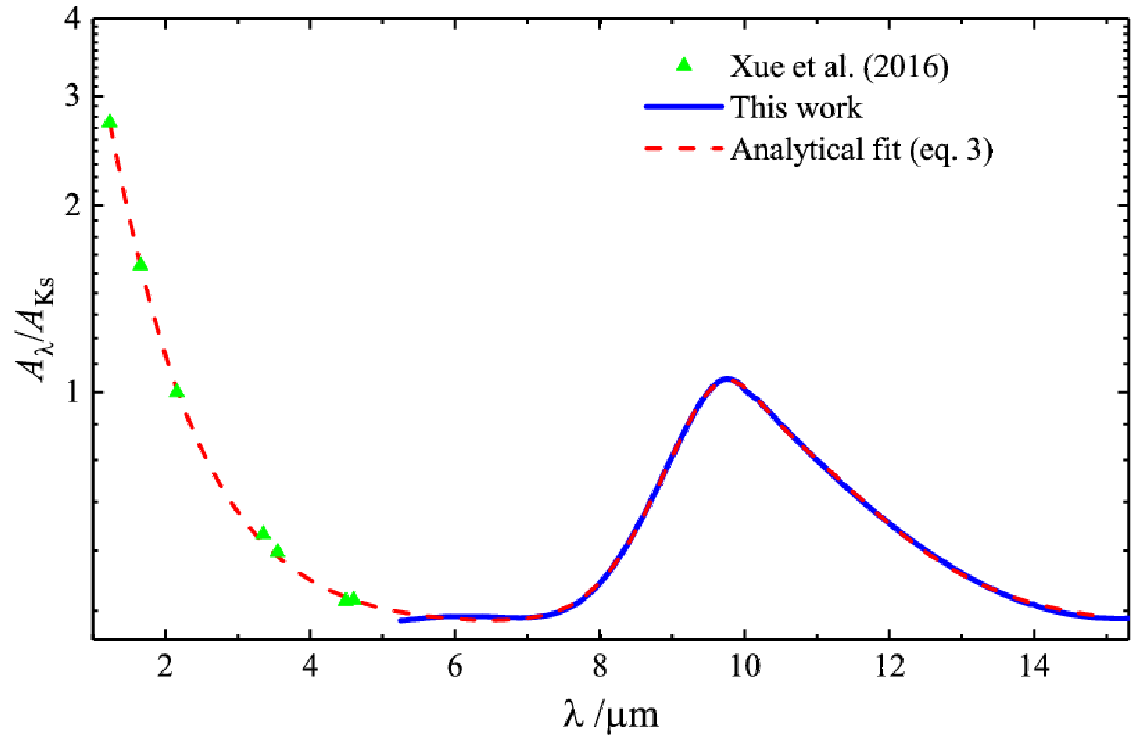}
\caption{\label{fig:polyextinc}
An analytical fit to the observed interstellar extinction curve in the IR at $0.9\simlt \lambda \simlt 15 \mum$.}
\end{figure*}

We fit the extinction in the near-IR
measured by Xue et al.\ (2016) and the extinction
around the 9.7$\mum$ silicate band
derived here with the following formula:
\begin{eqnarray}
\nonumber
A_\lambda/\AKs & =  & a_0
   + a_1\exp\left(-\lambda/\lambda_1\right)
   + a_2\exp\left(-\lambda/\lambda_2\right)  \\
\nonumber
&    & + \frac{a_3}
       {1 + \exp\left[-\left(\lambda-\lambda_3+0.5\,w_1\right)/w_2\right]}\\
\nonumber
&    & \times \left\{1-\frac{1}
       {1 + \exp\left[-\left(\lambda-\lambda_{3}-0.5\,w_1\right)/w_{3}\right]}\right\} \\
&    & + \frac{a_{4}}{w_{4}\sqrt{\pi/2}}
      \times\exp\left\{-2\left[\left(\lambda-\lambda_{4}\right)/w_{4}\right]^{2}\right\} ~~,
\end{eqnarray}
where $a_{0} =0.41980$;
$a_{1} =21.82920$,
$\lambda_{1} =0.42007\mum$;
$a_{2} =3.83330$,
$\lambda_{2} =1.02373\mum$;
$a_{3} =0.82771$,
$\lambda_{3} =9.87221\mum$,
$w_{1} =1.72679\mum$,
$w_{2} =0.44642\mum$,
$w_{3} =1.06985\mum$;
$a_{4} =0.13066$,
$\lambda_{4} =9.72506\mum$, and
$w_{4} =0.81006\mum$.
In Figure \ref{fig:polyextinc}
we show the observationally determined
IR extinction and the analytically fitted extinction.
We should note that eq.\,(3) is valid for
$0.9\mum\simlt\lambda\simlt15\mum$ and
should not be extrapolated to longer wavelengths
since $A_\lambda$ given by eq.\,(3) does not
seem to decline more rapidly than $1/\lambda$
which is required by the Kramers-Kronig relation
(see Draine 2004).

\section{Summary}
We have selected one O-type and
four B-type stars to trace
the interstellar silicate extinction.
For each star, we have determined
the silicate extinction profile around 9.7$\mum$
by comparing its \emph{Spitzer}/IRS spectrum
with that of unreddened reference star.
Our principal results are as follows:
\begin{enumerate}
\item The silicate extinction features of
          all our five sources peak around
          $\lambdapeak$\,$\simali$9.69--9.87$\mum$,
          appreciably longer than that of
          ``astronomical silicates''.
\item The width of the 9.7$\mum$ silicate
          extinction feature appears to
          bifurcate into two groups,
          a narrow one with a FWHM
          of $\gammasil\approx2.0\mum$ (for three stars)
          and a broad one with
          $\gammasil\approx3.0\mum$
          (for two stars).
          The width does not show any environmental dependence.
\item With $\lambdapeak\approx9.75\mum$
          and $\gammasil\approx2.2\mum$,
          the mean 9.7$\mum$ extinction profile,
          obtained by averaging over our five stars,
          closely resembles that of the diffuse interstellar
          medium along the lines of sight toward
          Cyg OB2 No.\,12 for which
          $\lambdapeak\approx9.74\mum$
          and $\gammasil\approx2.28\mum$
          and WR\,98a for which
          $\lambdapeak\approx9.77\mum$
          and $\gammasil\approx2.35\mum$.
\item The mean ratio of
          the visual extinction
          to the 9.7$\mum$ silicate absorption
          optical depth is
          $\langle \AV/\Delta\tau_{9.7}\rangle\approx18.2$,
          in close agreement with
          that of the solar neighborhood diffuse ISM
          but substantially exceeding
          that of the Galactic center
          and the dust torus around AGNs.
\item The color excess ratio at the peak wavelength
          of the 9.7$\mum$ silicate feature,
          $E(\lambdapeak-\Ks)/E(J-\Ks)$,
          is somewhat higher for the lines of sight
          with a higher CO (1-0) line intensity.
         We interpret this in terms of grain size effects:
         in dense regions, the presence of large grains
         would raise the continuum extinction underlying
         the 9.7$\mum$ feature and thus
         increase $E(\lambdapeak-\Ks)/E(J-\Ks)$.
\item An analytical formula is presented
          to parameterize the interstellar IR extinction.
          %from the  far ultraviolet to the mid-IR.
%
\end{enumerate}

%\vspace{-5mm}
\section*{Acknowledgements}
We thank B.T.~Draine, S.~Gao and
the anonymous referee for their very
helpful comments and suggestions.
%We thank the anonymous referee for
%his/her very instructive
%comments and suggestions.
This work is supported by
NSFC through Projects 11373015, 11533002, U1631104,
and 973 Program 2014CB845702.
This work made use of the data
taken mainly by \emph{Spitzer}/IRS, \emph{WISE} and 2MASS.

%%% References %%%

%\vspace{-5mm}

\clearpage

%********* Table 1: basic information *********
\begin{table}
%\tiny
%\scriptsize
\begin {center}
\caption{\label{table:t1}
Stellar parameters for the target and reference stars}
%\begin{tabular}{p{1.0cm}|p{0.7cm}|p{0.8cm}|p{0.8cm}|p{0.3cm}|p{0.7cm}|p{1cm}|p{0.8cm}}%{l|l|c|c|c|c|c|r}
\begin{tabular}{l|l|c|c|c|c|c|c|r}
\hline
\hline
Object & Type & $l$  & $b$
       & S/N & $J-\Ks$ & $Err(J-K_S)$ &$W_{\rm CO}$ & Note\\
       & & (degrees)& (degrees)& & (mag) & (mag) & (K\,km$\s^{-1}$)& \\
\hline
StRS 354 &  O7	&	076.97 & $-00.64$ & 57 & 1.88& 0.03&18.95& target \\
 %VI Cyg 11 & O5I & 080.5663 & +00.8336 & 59 & 0.660&7.418\\
 %VI Cyg 7 & O3I & 080.2426 & +00.8049 & 42 & 0.637&29.795\\
StRS 136 &   B8I	&	000.04 & $-00.57$ & 74 & 1.78& 0.03&425.26& target\\
StRS 164   &   B8I	&	014.21  & $-00.00$ & 69 & 1.78& 0.03&185.70& target\\
%BD+43 3710  &  B5I	&	083.3755 & +00.3400 & 87 & 0.717&16.047\\
HD 116119  & B8Ia & 306.62 & +00.63 & 59 & 0.42 & 0.02&40.85& target\\
%HD 112272 & B1Ia & 303.4864 & $-01.4947$ & 70 & 0.397&6.000\\
HD 147701  & B5III(?) & 352.25 & +16.85 & 43 & 0.48& 0.03&13.50& target\\
HD 204172  & B0I & 83.39 & $-$9.96 & 47 & $-$0.06 &0.03 & 0.03& reference\\
%HD 147701  & B5III(?) & 352.2547 & +16.8492 & 43 & 0.482&13.503\\
HD 128207 & B8V & 323.84 & +18.35 & 76 & $-$0.06& 0.03& 0.04 & reference\\
%27565312 &	IRAS 01342+6429  &  A6Iab	&  127.8530 & +02.3114 \\

\hline
\end{tabular}
\end{center}
%%\tablenotetext{}{'ref' means this star is a reference star with little even zero extinction}
\end{table}

%********* Table 1: basic information *********

%********* Table 2: photometry *********
%%% Table 2 %%%
%\begin{sidewaystable}[h]
\begin{landscape}
\begin{table}
%\begin{table}
\scriptsize
%\footnotesize
\begin {center}
\caption{\label{table:t2}
IR photometry for the target and reference stars }
\begin{tabular}{l|c|c|c|c|c|c|c|c|c|c|c|c|c|r}
%\tabletypesize{\tiny}
%\tablewidth{0truein}
%{\tiny
\hline
\hline
Object & J & H & $\rm \Ks$&W1& IRAC1 & IRAC2 & W2 & IRAC3 & IRAC4 & AKARI9 & W3 & AKARI18 & W4 & \\
       & 1.235$\mum$ &1.662$\mum$  &2.159$\mum$ & 3.353$\mum$ & 3.507$\mum$& 4.437$\mum$ &4.603$\mum$ &5.628$\mum$ &7.589$\mum$ &8.228$\mum$ &11.561$\mum$ &17.609$\mum$ &22.088$\mum$& Note\\
     & (mag)& (mag) & (mag) & (mag)& (mag) & (mag) & (mag) & (mag) & (mag) & (mag) & (mag) & (mag) & (mag) &  \\
\hline

StRS 354 &7.620 & 6.369 & 5.736 & 5.352& & & 4.847&  &  & 4.745 & 5.073&  &4.669 & target \\
%VI Cyg 11 & 6.650	&6.226	&5.990	&5.828 &  &  & 5.557 &  &  & 0.344 & 5.505 & & 4.870\\
%VI Cyg 7 & 7.248	&6.818	&6.611	&6.193 &  &  &  6.232 &  &  & 0.209 & 6.173 & &4.008:  \\
StRS 136 &6.880&5.725&5.101&4.338 & 4.685 &4.736 & 4.227&4.383 &4.363&4.318 &4.562 &3.824 &4.129 & target\\
StRS 164 & 7.662&6.499&5.886&5.262&  & 5.992& 4.993&5.212&5.245&5.198&5.527& &6.670:  & target\\
%BD+43 3710	&6.598&6.143&5.881&5.633&  &  &5.430&  &  &0.352 &5.486&  &4.195:  \\
HD 116119 & 6.121	&5.876	&5.697	&5.599 & 5.728 &5.548 & 5.351 & 5.518 & 5.478 & 5.307 & 5.488 & & 5.276& target\\
%HD 112272 & 5.573	&5.363	&5.176	&5.099 &  &  & 4.814 &  &  &  & 4.947 & & 4.651   \\
HD 147701 & 6.668	&6.378	&6.186	&6.046 &   &  & 5.956 &  &   &  &  6.137 &  &  5.886 & target \\
HD 204172 & 6.085& 6.080 & 6.141 &  6.140  &  &  &  6.056  &   &  &  6.001  &  6.041  &  &  5.736 & reference\\
%A6Iab &7.3&6.5& & 4.837& 4.431&4.088&3.874&3.339&3.787&3.62\\
%HD 147701 & 6.668	&6.378	&6.186	&6.046 &   &  & 5.956 &  &   &  &  6.137 &  &  5.886  \\
 HD 128207  &  5.959 & 6.052 & 6.015 & 6.048  &  &  &  5.992  &  &  &  5.807  &  6.087 & & 5.882& reference \\
\hline
\end{tabular}
\end{center}

%\tablenotetext{}{':' means the photometry in corresponding band is uncertain.}
\tiny  `:' The photometric data is somewhat uncertain.
%\end{table}
\end{table}
\end{landscape}
%\end{sidewaystable}

%********* Table 2: photometry *********

%********* Table 3: FWHM *********
%%  Table 4 %%
\begin{table}
%\tiny
%\scriptsize
\begin {center}
\caption{\label{table:FWHM}
The peak wavelengths ($\lambdapeak$),
FWHMs ($\gammasil$), strengths
($S_{9.7}$; see Figure~\ref{fig:extinccon})
of the 9.7$\mum$ silicate extinction profiles.
Also tabulated are the visual extinction ($\AV$),
the optical depth of the 9.7$\mum$ absorption
feature ($\Delta\tau_{9.7}$),
and $\AV/\Delta\tau_{9.7}$.}
\begin{tabular}{l|c|c|c|c|c|c|r}
\hline
\hline
Object & $\lambdapeak$ &  $\gammasil$
           & $S_{9.7}$ & $\AV$& $\Delta\tau_{9.7}$& $\AV/\Delta\tau_{9.7}$ & Reference \\
           & ($\mu$m) &  ($\mu$m) & &(mag)& & & \\
\hline

StRS 354 & 9.75&1.78 &0.33 &12.2&0.58& 21.0&\\
%VI Cyg 11 & 9.754& 2.10 \\
%VI Cyg 7 &  9.814& 1.85 \\
StRS 136 & 9.69&2.11&0.42 & 11.9&0.71& 16.8&\\
StRS 164 &   9.81&2.99&0.50&11.4&0.82&14.0 &\\
%BD+43 3710	& 9.75& 2.07\\
HD 147701 & 9.87&3.07 &0.35 &3.01& 0.17& 17.7& \\
HD 116119 &  9.75&1.90 & 0.25&2.38&0.11& 21.6&\\
%HD 112272 &  9.87&3.51\\
Average & 9.75 & 2.20 &0.36 &&&18.2 & \\
\hline
   WD01 $ (R_V=3.1)$ &  9.51  & 3.01& &&&14.5 & \\
    WD01 $ (R_V=5.5)$ & 9.45  & 3.55& & & & 14.2& \\
\hline
   WR 98a & 9.77& 2.35& &12.4&0.78& 15.9&  Chiar \& Tielens 2006\\
   Cyg OB2 No.\,12  & 9.74& 2.28& &10.2&0.59& 17.3& Fogerty et al. 2016\\
   GC Sgr A$^{\ast}$ & 9.77 & 1.73& &&& & Kemper et al. 2004\\
%\hline
%    Galactic center &    &  3.40 & & \citealp{Ro84,Ro85}\\
%    G, M supergiants and M giants  &   &   2.20 & & \citealp{Pegourie85}\\
%\hline
%      WC8/9 stars &  & 2.40 & & \citealp{Ro84,Ro85}\\
%      Sco and Galactic center stars  &   &  3.20 & &\citealp{Rieke85} \\
\hline
\end{tabular}
\end{center}
%\tablenotetext{}{
%$\rm {}^{a}$\citealp{Ro84,Ro85}\\
%$\rm {}^{b}$\citealp{Pegourie85}\\
%$\rm {}^{c}$\citealp{Rieke85}
%}
\end{table}
%********* Table 3: FWHM *********

%%% Table 4: V-Ks %%%
\begin{table}
%\tiny
%\scriptsize
\begin {center}
\caption{\label{table:VKs}
The $J-\Ks$ and $V-\Ks$ colors of
             our five source stars and the intrinsic
             $\left(J-\Ks\right)_0$ and
             $\left(V-\Ks\right)_0$ colors
             derived from the Kurucz model
             spectra for stars of the same spectral
             and luminosity type.
             Also tabulated are the color excesses
             $E(J-\Ks)$ and $E(V-\Ks)$.}
%\prescript{a}{}
\begin{tabular}{l|c|c|c|c|c|c|c}
\hline
Object & Type &$J-\Ks$$^a$
           & $(J-\Ks)_0$ & $E(J-\Ks)$
           & $V-\Ks$$^b$ & $(V-\Ks)_0$$^c$
           & $E(V-\Ks)$\\
 & & (mag)& (mag)& (mag)
     & (mag)& (mag)& (mag)\\
\hline
StRS 354& O7&1.884 &$-0.056$ &$ 1.94$ &$10.229$&	$-0.82^*$&$ 11.049$\\
StRS 136&B8I&1.779&$-0.056$ & $1.835$ & $10.899$ & $0.07$ &	10.829\\
StRS 164 & B8I&1.776 & $-0.056$ & $1.832$ & $10.414$ & 0.07 &10.344\\
HD 147701 &B5III&0.482 & $-0.056$ & $0.538$ & $2.281$ & $-0.42^{**}$ &2.701\\
HD 116119 &B8I&0.424 & $-0.056$ & $0.48$ & $2.173$ & $0.07$ & $2.103$ \\
\hline
\end{tabular}
\end{center}
%%\tablenotetext{}{$\lambda$ represents the color excess ratio $E(\lambda - \Ks)/E(J-\Ks)$ .
%}
\tiny a: Data taken from {\it 2MASS}\\
\tiny b: Data taken from {\it SIMBAD}\\
\tiny c: Data taken from Cox(2000)\\
\tiny *: Intrinsic color index of O9I
            due to the lack of data for O7I-type stars\\
\tiny **: Intrinsic color index of B5V
              due to the lack of data for B5III-type stars
\end{table}
%%% Table 4: V-Ks %%%

%********* Table 5: excess ratio *********
%%% Table 5 %%%
\begin{table}
%\tiny
%\scriptsize
\begin {center}
\caption{\label{table:t5}
              Color excess ratios
              $E(\lambda - \Ks)/E(J-\Ks)$
              for the {\it 2MASS} H band
              and the {\it WISE} W1,W2,W3
              and W4 bands and at $\lambdapeak$,
              the peak wavelength of the 9.7$\mum$
              silicate extinction profiles
              determined in this work.
              Also shown are the model ratios
              of WD01 for $R_V=3.1$
              and $R_V=5.5$ as well as
              that of Xue et al. (2016).
              }
\begin{tabular}{l|c|c|c|c|c|r}
\hline
\diagbox{Star}{Band} & H & W1 & W2 & W3 & W4 & $\lambdapeak$\\
\hline

StRS 354& 0.358 &$-0.197$ &$ -0.414$ &$-0.290$&	$-0.341$&$	-0.073$\\
%VI Cyg 11& 0.289&-0.188& -0.435&	-0.579&-1.108&-0.272\\
%VI Cyg 7&0.258&-0.545&	-0.375&	-0.533&	-3.18:&-0.206\\
StRS 136&0.373&$-0.415$ & $-0.430$ & $-0.239$ & $-0.309$ &	0.076\\
StRS 164 & 0.368 & $-0.340$ & $-0.441$ & $-0.141$ & 0.649: &0.181\\
%BD+43 3710 &0.418 &-0.320&-0.473&	-0.382&-1.657:&	-0.163\\
HD 147701 &0.288 & $-0.322$ & $-0.385$ & $-0.225$ & $-0.310$ &0.040\\
HD 116119 &0.500 & $-0.202$ & $-0.544$ & $-0.227$ & $-0.033$ & $-0.069$ \\
%HD 112272 &0.547&-0.168 &-0.552&-0.285&	-0.265&-0.016\\
Average & 0.377& $-0.295$ & $-0.443$ & $-0.224$ & $-0.248$ &0.031\\
RMS & 0.077&0.094&0.060&0.054&0.144&0.107\\
\hline
WD01 $R_V=3.1$&0.392& $-0.359$ & $-0.494$ & $-0.410$ & $-0.531$ & $-0.162$ \\
WD01 $R_V=5.5$ &0.331 & $-0.248$ & $-0.354$ & $-0.309$ & $-0.501$ & $-0.008$ \\
\hline
Xue et al.\ (2016) & 0.348& $-0.238$ & $-0.312$ & $-0.269$ & $-0.370$ & \\

%A6Iab &7.3&6.5& & 4.837& 4.431&4.088&3.874&3.339&3.787&3.62\\

\hline
\end{tabular}
\end{center}
%%\tablenotetext{}{$\lambda$ represents the color excess ratio $E(\lambda - \Ks)/E(J-\Ks)$ .
%}
\end{table}

%********* Table 5: excess ratio *********

%********* Figure 1: otype *********
%********* Figure 1: otype *********

%********* Figure 2: btype *********
%********* Figure 2: btype *********

%********* Figure 3: refbijiao *********
%********* Figure 3: refbijiao *********

%********* Figure 4: refduibi *********
%********* Figure 4: refduibi *********

%********* Figure 5: sourcenihe *********
%********* Figure 5: sourcenihe *********

%********* Figure 6: extinccon *********
%********* Figure 6: extinccon *********

%*** Figure 7: Normalized extinction profile ***
%*** Figure 7: Normalized extinction profile ***

%********* Figure 8: obme *********
%********* Figure 8: obme *********

%********* Figure 9: obdai *********
%********* Figure 9: obdai *********

%********* Figure 10: profilevs *********
%********* Figure 10: profilevs *********

%********* Figure 11: profilevs *********
%********* Figure 11: profilevs *********

%********* Figure 12: kw3jk *********
%********* Figure 12: kw3jk *********

%********* Figure 13: polyextinc *********
%********* Figure 13: polyextinc *********

%\bsp
\label{lastpage}


\begin{thebibliography}{}
\bibitem[Barlow \& Cohen (1977)]{Barlow77}Barlow, M. J., \& Cohen, M. 1977, ApJ, 213, 737
%\bibitem[Beichiman et al.(1988)]{Beichiman88}Beichman, C. A., Neugebauer, G., Habing, H. J., Clegg, P. E., \& Chester, T. J., ed. 1988, Infrared Astronomical Satellite (IRAS) Catalogs and Atlases.Volume 1: Explanatory Supplement, Vol. 1
\bibitem[Bohren \& Huffman (1983)]{Bohren83}Bohren, C.F., \& Huffman, D. R. 1983, Absorption and Scattering of Light by Small Particles,
Wiley, New York
\bibitem[Cardelli et al.(1989)]{Cardelli89}Cardelli, J. A., Clayton, G. C.,
                  \& Mathis, J. S.\ 1989, ApJ, 345, 245 (CCM)
\bibitem[Castelli et al.(1997)]{Castelli97}Castelli, F., Gratton, R. G., \& Kurucz, R. L. 1997, A\&A, 318, 841
\bibitem[Chen et al.(2016)]{Chen16} Chen, R., Luo, A., Liu, J. M., \& Jiang, B.W. 2016, AJ, 151, 146
\bibitem[Chiar \& Tielens (2006)]{Chiar06} Chiar, J. E., \& Tielens, A. G. G. M. 2006, ApJ, 637, 774
%\bibitem[Cutri et al.(2003)]{Cutri03} Cutri, R. M., et al. 2003, 2MASS All Sky Catalog of point sources.
%¡ª. 2013, Explanatory Supplement to the AllWISE Data Release Products, Tech. rep.
\bibitem[Cox 2000]{cox2000}Cox, A. N. 2000, ed., Allen¡¯s Astrophysical Quantities, Springer, New York
\bibitem[Davenport et al.(2014)]{Davenport14}Davenport, J. R. A., Ivezi$\rm \acute{c}$, $\rm \check{Z}$., Becker, A. C., et al. 2014, MNRAS, 440, 3430
\bibitem[Dorschner et al.(1995)]{Dorschner95}Dorschner, J., Begemann,
  B., Henning, Th., Jaeger, C., \& Mutschke, H. 1995, A\&A, 300, 503
\bibitem[Draine (1984)]{Draine1984}Draine, B.~T.\ 1984, ApJL, 277, L71
\bibitem[Draine (1989)]{Draine1989}Draine, B. T. 1989, in Infrared Spectroscopy in Astronomy, ed. B. H. Kaldeich (ESA Special Publication, Vol. 290; Noordwijk: ESA), 93
\bibitem[Draine (2003)]{Draine03}Draine, B. T. 2003, ARA\&A, 41, 241
\bibitem[Draine (2004)]{Draine04}Draine, B. T. 2004, in The Cold Universe, ed. A. W. Blain et al. (Berlin:
Springer), 213
\bibitem[Draine (2016)]{Draine2016}Draine, B.~T.\ 2016, ApJ, 831, 109
\bibitem[Draine \& Lee (1984)]{Dr84} Draine, B.T., \& Lee, H.M. 1984, ApJ, 318, 485
%\bibitem[Draine \& Shapiro (1989)]{Dr89}Draine, B.T., \& Shapiro, P.R. 1989, Ap. J. Lett., 344, L45-L48.
%\bibitem[Ducati et al.(2001)]{Ducati01} Ducati, J. R., Bevilacqua, C. M., Rembold, S. B., \& Ribeiro, D. 2001, ApJ, 558, 309
\bibitem[Flaherty et al.(2007)]{Flaherty07} Flaherty, K. M., Pipher, J. L., Megeath, S. T., et al. 2007, ApJ, 663, 1069
\bibitem[Fogerty et al.(2016)]{Fogerty16} Fogerty, S., Forrest, W., Watson, D. M., Sargent, B. A., \& Koch, I. 2016, ApJ, 183, 71
\bibitem[Forrest et al.(1975)]{Forrest75}Forrest, W. J., Gillett, F. C., \& Stein, W. A. 1975, ApJ, 195, 423
\bibitem[Gao et al.(2009)]{Gao09} Gao, J., Jiang, B. W., \& Li, A. 2009, ApJ, 707, 89
%\bibitem[Gehrz et al.(1974)]{Gehrz74}Gehrz R. D., Hackwell J. A., \& Jones T. W., 1974, ApJ, 191, 675
\bibitem[Gillett et al.(1975)]{Gillet75}Gillett, F. C., Forrest, W. J., Merrill, K. M., Soifer, B. T., \& Capps, R. W. 1975, ApJ, 200, 609
\bibitem[Henning (2010)]{Henning10} Henning, Th., 2010, ARA\&A, 48, 21
\bibitem[Henning \& Meeus (2011)]{Henning11}Henning, Th.,
              \& Meeus, G. 2011, in Physical Processes
              in Circumstellar Disks around Young Stars,
              ed. P. J. V. Garcia
              (Chicago, IL: Univ. Chicago Press), 114
%\bibitem[Houck et al.(2004)]{Houck04} Houck, J. R., et al. 2004, ApJS, 154, 18
\bibitem[Indebetouw et al.(2005)]{Indebetouw05} Indebetouw, R.,
               Mathis, J. S., Babler, B. L., et al.\ 2005, ApJ, 619, 931
\bibitem[J\"ager et al.(1994)]{Jager94}J\"aeger, C.,
              Mutschke, H., Begemann, B., Dorschner, J.,
              \& Henning, Th. 1994, A\&A, 292, 641
\bibitem[J\"ager et al.(2003)]{Jager03} J\"ager, C.,
              Dorschner, J., Mutschke, H., Posch, T.,
              \& Henning, Th. 2003, A\&A, 408, 193
\bibitem[Jarrett et al.(2011)]{Jarrett11}Jarrett, T. H., Cohen, M., Masci, F., et al. 2011, ApJ, 735, 112
%\bibitem[Jiang et al.(2003)]{Jiang03} Jiang, B. W., Omont, A., Ganesh, S., Simon, G., \& Schuller, F. 2003, A\&A, 400, 903
\bibitem[Kemper et al.(2004)]{Kemper04} Kemper, F., Vriend, W. J., \& Tielens, A. G. G. M. 2004, ApJ, 609, 826
\bibitem[Koornneef (1983)]{Koornneef83}Koornneef, J., 1983, A\&A, 128, 84
\bibitem[Lebouteiller et al.(2011)]{Lebouteiller11}Lebouteiller, V., Barry, D. J., Spoon, H. W. W., et al. 2011, ApJS, 196, 8
\bibitem[Li \& Draine (2001)]{Li01}Li, A., \& Draine, B. T. 2001, ApJ, 550, L213
\bibitem[Li et al.(2015)]{Li2015}Li, A., Wang, S., Gao, J.,
              \& Jiang, B.W.\ 2015,
              in Lessons from the Local Group,
              ed. Freeman, K.C., Elmegreen, B.G.,
              Block, D.L. \& Woolway, M.
              (Springer: New York), 85 (arXiv: 1507.06604)
\bibitem[Li et al.(2008)]{Li08}Li, M. P., Shi, Q. J., \& Li, A. 2008, MNRAS, 391, L49
%\bibitem[Lutz (1999)]{Lutz99} Lutz, D. 1999, in ESA Special Publication, Vol. 427, The Universe as Seen by ISO, ed. P. Cox
%\& M. Kessler, 623
%\bibitem[Lutz et al.(1996)]{Lutz96} Lutz, D., et al. 1996, A\&A, 315, L269
\bibitem[Majewski et al.(2011)]{Majewski11} Majewski, S. R., Zasowski, G., Nidever, \& D. L. 2011, ApJ, 739, 25
\bibitem[Martin \& Whittet (1990)]{Martin90} Martin, P. G., \& Whittet, D. C. B. 1990, ApJ, 357, 113
\bibitem[McClure (2009)]{McClure09}McClure, M. 2009, ApJ, 693, L81
%\bibitem[Meyer et al.(2007)]{Meyer07}  Meyer, M. R., Backman, D. E., Weinberger, A. J., \& Wyatt, M. C. 2007, in
%Protostars and Planets V, ed. B. Reipurth, D. Jewitt \& K. Keil (Tucson, AZ:Univ. Arizona Press), 573
\bibitem[Olofsson \& Olofsson (2011)]{Olofsson} Olofsson, S. \& Olofsson, G. 2011, A\&A, 534, 127
\bibitem[Pegourie \& Papoular(1985)]{Pegourie85} Pegourie, B., \& Papoular, R. 1985, A\&A, 142, 451
\bibitem[Poteet et al.(2015)]{Poteet15}Poteet, C.A., Whittet, D.C.B., \& Draine, B.T. 2015, ApJ, 801, 110
\bibitem[Planck Collaboration XIII(2014)]{Planck14} Planck Collaboration XIII. 2014, A\&A, 571, A13
\bibitem[Rice et al.(2016)]{Rice16}Rice, T.S., Goodman, A.A., Bergin, E.A., Beaumont, C., \& Dame, T.M. 2016, ApJ, 822, 52
\bibitem[Rieke \& Lebofsky (1985)]{Rieke85} Rieke, G. H., \& Lebofsky, M. J. 1985, ApJ, 288, 618
\bibitem[Roche \& Aitken.(1984)]{Ro84} Roche, P. F., \& Aitken, D.K. 1984, MNRAS, 208, 481
\bibitem[Roche \& Aitken.(1985)]{Ro85} Roche, P. F., \& Aitken, D.K. 1985, MNRAS, 215, 425
\bibitem[Rouleau \& Martin (1991)]{1991ApJ...377..526R} Rouleau, F., \& Martin, P.~G.\ 1991, ApJ, 377, 526
\bibitem[Russell et al.(1975)]{Russell75}Russell, R. W., Soifer, B. T., \& Forrest, W. J. 1975, ApJL, 198, 41
\bibitem[Seok \& Li (2017)]{Seok17}Seok, J. Y., \& Li, A. 2017, ApJ, 835, 291
\bibitem[Shao et al.(2017)]{Shao17} Shao, Z. Z., Jiang, B. W., \& Li, A. 2017, ApJ, 840, 27
\bibitem[Sjouwerman et al.(2009)]{Sjouwerman09}Sjouwerman, L. O., Capen, S. M., \& Claussen, M. J. 2009, ApJ, 705, 1554
\bibitem[van Breemen et al.(2011)]{Breemen11} van Breemen, J. M., Min, M., Chiar, J.E., et al. 2011 A\&A 526, A152
\bibitem[van der Veen \& Habing (1988)]{vanderveen88}van der Veen, W. E. C. J., \& Habing, H. J. 1988, A\&A, 194, 125
\bibitem[Wang et al.(2013)]{Wang13}Wang, S., Gao, J., Jiang, B. W., Li, A., \& Chen, Y. 2013, ApJ, 773, 30
\bibitem[Wang \& Jiang (2014)]{Wang14} Wang, S., \& Jiang, B. W. 2014, ApJ, 788, L12
%\bibitem[Wang et al.(2015)]{Wang15}Wang, S., Li, A., \& Jiang, B. W. 2015, MNRAS, 454, 569
\bibitem[Weingartner \& Draine (2001)] {WD01} Weingartner, J. C.,
              \& Draine, B. T. 2001, ApJ, 548, 296 (WD01)
%\bibitem[Werner et al.(2004)]{Werner04} Werner, M. W., Roellig, T. L., Low, F. J., et al. 2004, ApJS, 154, 1
%\bibitem[Whittet \& van Breda (1980)]{Whittet80}Whittet, D. C. B., \& van Breda, I. G. 1980, MNRAS, 192, 467
\bibitem[Whittet (2003)]{whittet03}Whittet, D. C. B.\ 2003, Dust in the Galactic Environment (2nd ed;
Bristol: Institute of Physics Publishing)
%\bibitem[Wright et al.(2010)]{Wright10} Wright, E. L., et al. 2010, AJ, 140, 1868
\bibitem[Xue et al.(2016)]{Xue16} Xue, M. Y.,
              Jiang, B. W., Gao, J., et al.\ 2016, ApJS, 224, 18
\bibitem[Yuan et al.(2013)]{Yuan13} Yuan, H. B., Liu, X. W., \& Xiang, M. S. 2013, MNRAS, 430, 2188
\bibitem[Zasowski et al.(2009)]{Zasowski09} Zasowski, G., Majewski, S. R., Indebetouw, R., et al. 2009, ApJ, 707, 510

%
\end{thebibliography}
\end{document}